\documentclass[reprint,amsmath,amssymb,aps,prx,superscriptaddress,longbibliography,floatfix]{revtex4-2}
\usepackage{graphicx}
\usepackage[utf8]{inputenc}
\usepackage{soul}
\usepackage[usenames]{xcolor}
\usepackage{dcolumn}
\usepackage{bm}
\usepackage{textgreek}  

\DeclareMathOperator*{\argmax}{arg\,max}

\newcommand{\TMCT}{T_\mathrm{MCT}}

\newcommand{\VIS}{V_\mathrm{IS}}
\newcommand{\xiPTS}{\xi_\mathrm{PTS}}
\newcommand{\Tf}{T_\mathrm{f}}

\begin{document}

\title{Dynamical Facilitation Governs the Equilibration Dynamics of Glasses}

\author{Rahul N. Chacko}
\affiliation{Department of Physics and Astronomy, University of Pennsylvania, Philadelphia, Pennsylvania 19104, USA}

\author{Fran\c{c}ois P. Landes}
\affiliation{Universit\'{e} Paris-Saclay, CNRS, INRIA, Laboratoire Interdisciplinaire des Sciences du Num\'{e}rique, TAU team,  91190 Gif-sur-Yvette, France}

\author{Giulio Biroli}
\affiliation{Laboratoire de Physique de l'\'{E}cole Normale Sup\'{e}rieure, ENS, Universit\'{e} PSL, CNRS, Sorbonne Universit\'{e},
Universit\'{e} Paris-Diderot, Sorbonne Paris Cit\'{e}, 75005 Paris, France}

\author{Olivier Dauchot}
\affiliation{UMR Gulliver 7083 CNRS, ESPCI, PSL Research University, 10 rue Vauquelin, 75005 Paris, France}

\author{Andrea J. Liu}
\affiliation{Department of Physics and Astronomy, University of Pennsylvania, Philadelphia, Pennsylvania 19104, USA}

\author{David R. Reichman}
\affiliation{Department of Chemistry, Columbia University, 3000 Broadway, New York, New York 10027, USA}

\date{\today}

\begin{abstract}
Convincing evidence of domain growth in the heating of ultrastable glasses
suggests that the equilibration dynamics of super-cooled liquids could be driven by a nucleation and growth mechanism.
We investigate this possibility by simulating the equilibration dynamics of a model glass
during both heating and cooling between poorly and well-annealed states.
Though we do observe the growth of domains during heating,
we find that domains are absent during cooling.
This absence is inconsistent with classical nucleation theory.
By comparing the equilibration dynamics of our glass
with that of two models with kinetic constraints,  
we demonstrate that dynamical facilitation
generically leads to heating driven by domain growth and cooling without domains.
Our results provide strong evidence that dynamical facilitation, not nucleation and interfacial-tension-driven domain growth, is the driving mechanism for the equilibration dynamics of glass-formers.
\end{abstract}

\maketitle

%
%

\section{Introduction \label{sec:intro}}

The phenomenology of glasses and supercooled liquids
(\textit{i.e.}, glasses in metastable equilibrium such that detailed balance is satisfied on all relevant time scales)
bears a striking resemblance to that of crystalline solids%
~\cite{kirkpatrickScalingConceptsDynamics1989, douglassCanStableGlass2013, berthierEvidenceDisorderedCritical2015,
jackMeltingStableGlasses2016, cubetaGlassSofteningKinetics2019, vila-costaNucleationGrowthSupercooled2020,
herreroTwostepDevitrificationUltrastable2023, herreroFrontPropagationUltrastable2023}.
As with transformations of Eshelby disclinations in crystals~\cite{eshelbyDeterminationElasticField1957},
local plastic events in these systems perturb their neighborhoods with long-ranged quadrupolar displacement fields%
~\cite{picardElasticConsequencesSingle2004, chackoElastoplasticityMediatesDynamical2021}.
A large body of experimental evidence also finds that ultrastable glasses
evolve upon heating in a manner consistent with the Avrami framework of classical nucleation theory%
~\cite{fanfoniJohnsonMehlAvramiKohnogorovModelBrief1998}:
fronts of high-temperature supercooled liquid advance at constant speed
into the bulk ultrastable glass%
~\cite{swallenStableGlassTransformation2009, kearnsObservationLowHeat2010, kearnsOneMicrometerLength2010,
dawsonAnisotropicStructureTransformation2011, whitakerVapordepositedVtrisnaphthylbenzeneGlasses2012,
sepulvedaAnomalousTransformationVaporDeposited2012, chenDynamicsGlassformingLiquids2013,
sepulvedaManipulatingPropertiesStable2013, sepulvedaRoleFragilityFormation2014, rodriguez-tinocoEvaluationGrowthFront2014,
bhattacharyaEnthalpyHighTemperature2014, rodriguez-tinocoTransformationKineticsVapordeposited2015,
dalalInfluenceSubstrateTemperature2015, tylinskiVapordepositedGlassesMethyl2015,
waltersThermalStabilityVapordeposited2015, rafols-ribeRoleThermodynamicStability2017,
cubetaCommunicationSurfacefacilitatedSoftening2017, cubetaGlassSofteningKinetics2019},
with the fraction of material retaining ultrastable glass structure
decaying with time according to a compressed exponential law%
~\cite{kearnsOneMicrometerLength2010, dawsonAnisotropicStructureTransformation2011,
sepulvedaAnomalousTransformationVaporDeposited2012, fullertonDensityControlsKinetic2017}.
By contrast, poorly-annealed (``ordinary'') glasses are found to transform homogeneously into
the supercooled liquid upon heating above the glass transition temperature%
~\cite{sepulvedaManipulatingPropertiesStable2013, waltersThermalStabilityVapordeposited2015}.
This difference of behavior, also observed in pressure-controlled molecular dynamics (MD) simulations%
~\cite{fullertonDensityControlsKinetic2017, flennerFrontMediatedMeltingIsotropic2019,
herreroTwostepDevitrificationUltrastable2023, herreroFrontPropagationUltrastable2023},
suggests a process analogous to crystal freezing and melting%
~\cite{vila-costaNucleationGrowthSupercooled2020, ruiz-ruizRealtimeMicroscopyRelaxation2023,
herreroTwostepDevitrificationUltrastable2023, herreroFrontPropagationUltrastable2023}
and raises the possibility that a thermodynamic first-order transition
mediated by nucleation and domain growth separates poorly-annealed glass from well-annealed glass%
~\cite{angell_uncertain_2015, vila-costaNucleationGrowthSupercooled2020, herreroFrontPropagationUltrastable2023}.

Indeed, as with crystals, a notion of structural order also applies to supercooled liquids.
In these amorphous systems, local equilibrium structure constrains%
~\cite{dasSoftPinningExperimental2023} and is constrained by~\cite{berthierZerotemperatureGlassTransition2019}
neighboring structure up to a finite distance $\xiPTS$ (the ``point-to-set'' length) away~\cite{biroliRandomFirstOrder2012}.
Random First Order Transition Theory (RFOT), known to be exact in infinite dimension~\cite{charbonneauExactTheoryDense2014},
predicts systems below a critical temperature $\TMCT$ to be separated into a mosaic of
$\xiPTS$-sized domains of mutually-incompatible amorphous order%
~\cite{kirkpatrickScalingConceptsDynamics1989, bouchaudAdamGibbsKirkpatrickThirumalaiWolynesScenarioViscosity2004, biroliRandomFirstOrder2012}.
As demonstrated by randomly pinned glasses~\cite{hockyEquilibriumUltrastableGlasses2014}
and supercooled liquids coupled to a reference configuration~\cite{berthierEvidenceDisorderedCritical2015},
amorphous order can in principle lead to first order transitions in amorphous systems,
supporting the possibility of a first-order transition in supercooled liquids \cite{cammarotaGeneralApproachSystems2013}.

Nucleation and domain growth is not the only available explanation for the equilibration dynamics of glass-formers, however.
Particles in supercooled liquids are trapped into ``cages'' by their nearest neighbors,
rattling around for long periods between cage-escape events
in which they hop out of one cage and into another%
~\cite{weeksThreeDimensionalDirectImaging2000a, candelierSpatiotemporalHierarchyRelaxation2010}.
As a local plastic event within an otherwise elastic material,
a cage escape weakens its neighborhood, creating the conditions for more cage escapes%
~\cite{chackoElastoplasticityMediatesDynamical2021}.
Such a process, in which rearrangements propagate mobility to their local neighborhoods,
is known as dynamical facilitation~\cite{keysExcitationsAreLocalized2011, tarjusOverviewTheoriesGlass2011}
and plays an important role in glassy liquids \cite{dauchotDynamicalHeterogeneitiesGrains2011}.
This self-propagation of mobility
results in avalanches of plastic events correlated in space and time~\cite{candelierSpatiotemporalHierarchyRelaxation2010}.
On long time scales, these avalanches appear as large clusters of spatially-correlated mobility
that grow as a function of time scale~\cite{keysExcitationsAreLocalized2011, scallietThirtyMillisecondsLife2022},
resembling domain growth.
Dynamical facilitation plays an important role in the equilibrium dynamics of supercooled liquids,
and it could also be the key mechanism at play in {\it equilibration dynamics}, i.e. relaxation towards equilibrium. 
Indeed, many of the phenomena observed in glasses upon heating or cooling are also seen in Kinetically Constrained Models (KCMs),
spin lattice models in which spins interact only via facilitation~\cite{garrahanKineticallyConstrainedModels2011}, upon heating%
~\cite{butlerGlassyRelaxationSurfaces1991, douglassCanStableGlass2013, gutierrezFrontPropagationBulk2016}
or cooling~\cite{garrahanGlassinessConstrainedDynamics2000}.
Dynamical facilitation therefore offers a possible alternative explanation for
the behavior of glasses during heating~\cite{%
leonardMacroscopicFacilitationGlassy2010, sepulvedaManipulatingPropertiesStable2013}
and cooling~\cite{garrahanGlassinessConstrainedDynamics2000}.
In this paper, we show that it is indeed dynamical facilitation that governs the equilibration dynamics of glasses,
not nucleation followed by domain growth.

We first address the question of nucleation,
interpreted in the classical sense~\cite{kashchievNucleationBasicTheory2000}  
of a competition between bulk and interfacial free energies which sets a critical size
above which randomly-nucleated domains will grow and below which they will shrink.
We distinguish nucleation,
which involves a thermodynamic drive towards domain growth as small domains
(of linear size smaller than the nucleation length) are suppressed,
from the evolution of structure via random, thermally-activated rearrangements,
with no suppression of small domains.
In the latter case, an additional mechanism is required to explain domain growth.
In the second part of this paper, we show that dynamical facilitation fills this role.

We conduct MD simulations of a polydisperse glass
at fixed temperature $T_\mathrm{eq}$ and pressure $P$ for two types of initial condition.
In simulation runs using the first type of initial condition,
which we refer to as \emph{homogeneous simulations},
the initial state corresponds to equilibrium at an initial temperature $T_0$ different from the thermostat temperature $T_\mathrm{eq}$.
In simulation runs using the second type of initial condition,
which we refer to as \emph{slab simulations},
we introduce a system-percolating slab of $T_\mathrm{eq}$
equilibrium structure lying in the $x \equiv 0$ plane that interrupts the $T_0$ structure of the surrounding system
(see Figs.~\ref{fig:inhom_poly}(a) and \ref{fig:inhom_poly}(b)).
In this geometry, the slab can expand while keeping the size and shape of its boundary fixed,
avoiding any putative interfacial free energy cost as it expands.
In a classical nucleation theory scenario,
the slab represents a nucleus above the critical nucleus size.
We fix $T_\mathrm{low} = \min \left\{ T_0, T_\mathrm{eq} \right\}$ and $T_\mathrm{high} = \max \left\{ T_0, T_\mathrm{eq} \right\}$ so that
any critical point separating $T_0$ and $T_\mathrm{eq}$ during heating will also do so during cooling.

These simulations provide a number of tests for nucleation and domain growth:
\begin{enumerate}
    \item We can characterize the local structure and dynamics, 
    and simply look (by eye) for growing domains of $T_\mathrm{eq}$ structure in individual realizations
    of the homogeneous simulation.
    \item We can also look for evidence of a nucleation time in these simulations.
    \item We can quantitatively track domain growth via the displacement of the slab boundary in slab simulations.
    \item We can assess the roughness of domains, both qualitatively in individual realizations of the homogeneous simulation
    and quantitatively (via the width of the slab boundary) in the slab simulation.
\end{enumerate}
As we will show, all four tests argue against nucleation,
with the most striking evidence provided by the absence of domain growth during cooling, as we shall discuss.

We validate these tests by first applying them to a crystallizing monodisperse system with
temperatures $T_\mathrm{low}$ and $T_\mathrm{high}$ straddling its melting point,
verifying our expectations for the case of a genuine first order transition.
We then apply this methodology to our polydisperse model glass,
obtaining results inconsistent with the nucleation and growth mechanism.
Finally,
we support our claim that dynamical facilitation is responsible for the equilibration dynamics of glasses
by applying the same four tests to two plaquette models, the Triangular Plaquette Model (TPM)%
~\cite{garrahanGlassinessConstrainedDynamics2000, garrahanGlassinessEmergenceEffective2002, garrahanTransitionCoupledReplicas2014}
and the Square Pyramidal Plaquette Model (SPPM)~\cite{turnerOverlapActivityGlass2015}.

Plaquette models are spin lattice models with a dual representation in terms of plaquettes
with trivial thermodynamics and kinetically-constrained dynamics~\cite{ritortGlassyDynamicsKinetically2003},
corresponding to KCMs in the plaquette representation.
The TPM and SPPM are similar by design, and the mechanisms underlying dynamical facilitation in these two systems
differ greatly from the elastoplastic mechanism for facilitation in supercooled liquids and glasses%
~\cite{chackoElastoplasticityMediatesDynamical2021}.
We find that just like our polydisperse model glass,
these plaquette models exhibit domain growth during heating and none during cooling.
For all three systems,
domain growth during heating proceeds at a constant speed that persists even as the system relaxes.
The only notable difference between the systems is the evolution of domain roughness,
which is different for all three systems
and therefore not a test for facilitation-driven dynamics.

In investigating nucleation and domain growth and the role of dynamical facilitation as above,
we also establish two side-results of particular interest.
Firstly, we provide direct support for a two-state equilibration scenario in which glasses evolve from
their initial state into the target equilibrium state (set by the thermostat)
via direct transformation of structure into the target equilibrium structure,
as opposed to via one or more intermediate structural states.
This scenario has long been the experimental consensus for heating%
~\cite{kearnsObservationLowHeat2010, rodriguez-tinocoEvaluationGrowthFront2014,
rodriguez-tinocoTransformationKineticsVapordeposited2015, rafols-ribeRoleThermodynamicStability2017,
cubetaCommunicationSurfacefacilitatedSoftening2017, rodriguez-tinocoSurfaceBulkInterplayVaporDeposited2019,
vila-costaNucleationGrowthSupercooled2020,
vila-costa_emergence_2023}.
However, we notably also find it to hold for \emph{cooling}.
Secondly, to support dynamical facilitation as the governing mechanism for the equilibration dynamics of glasses,
we show that the equilibration dynamics of the TPM in the slab geometry,
which closely resembles that of the supercooled liquid,
can be captured using the same phenomenological model for both heating and cooling.

The structure of this paper is as follows.
In \S\ref{sec:systems}, we discuss in detail our model crystal and glass systems,
and the way in which our simulation protocols are applied to them.
In \ref{sec:crystal} and \ref{sec:glasses},
we apply our four tests to the model crystal and glass systems.
We defer a broader discussion of the significance of these results until \S\ref{sec:discussion}.
In \S\ref{sec:plaquette}, we introduce our two plaquette models
and show the results of applying the four tests to them.
Finally, in \S\ref{sec:discussion},
we synthesize the results for the four different systems,
discussing the conclusions that can be drawn from comparing these results with each other.
Readers may find it helpful to refer to \S VI after they read about each of the 4 model systems
(two particulate systems and two plaquette models) in turn.

\section{System \label{sec:systems}}

Our three-dimensional monodisperse crystal and two-dimensional polydisperse glass systems
comprise particles of uniform mass $m$ interacting via the pair potential
\begin{equation}
    V \left(\tilde r \right) = 
    \begin{cases}
        V_0 \left( \tilde r^{-12} + c_0 + c_2 \tilde r^2 + c_4 \tilde r^4 \right) & \tilde r \leq 1.25, \\
        0 & \tilde r > 1.25,
    \end{cases} \label{eq:pair_potential}
\end{equation}
where
\begin{equation*}
    \tilde r = 2 r /  \left [ \left( \sigma_i + \sigma_j \right) \left( 1 - 0.2 \left| \sigma_i - \sigma_j \right| \right) \right].
\end{equation*}
Here, $r$ is the separation between particles $i$ and $j$ with sizes $\sigma_i$ and $\sigma_j$,
and the constants $c_0$, $c_2$, and $c_4$ are chosen such that
$V \left(1.25 \right) = V^\prime \left( 1.25 \right) = V^{\prime \prime} \left( 1.25 \right) = 0$
(\S\ref{app:swap}).
Following \cite{ninarelloModelsAlgorithmsNext2017},
for the polydisperse system, $\sigma_i$ for each particle is drawn independently
from a distribution with probability density $\propto \sigma^{-3}$
in the interval $\left[ \sigma_\mathrm{min}, \sigma_\mathrm{max} \right]$,
with $\sigma_\mathrm{min}$ and $\sigma_\mathrm{max}$ chosen such that the coefficient of variation of $\sigma$
is $c_\sigma = 0.23$.
This system, introduced in \cite{ninarelloModelsAlgorithmsNext2017},
is designed to allow equilibration to temperatures well below $\TMCT$ using swap Monte Carlo simulations.
We take $V_0$, $\overline \sigma$, $m$, and $V_0 / k_\mathrm{B}$ to be our energy, length, mass, and temperature units,
where $k_\mathrm{B}$ is the Boltzmann constant and
where $\overline \sigma$ is the expected value of the particle diameter distribution.
The system coordinates are centered at the origin
and periodic boundary conditions are imposed along each axis of the system.

We prepare our initial states using constant-volume Monte Carlo simulations
implemented as described in \S\ref{app:prep}.
For the monodisperse case, the homogeneous simulations have $N=27648$ particles and initial lengths
$48$, $12 \sqrt{3}$, and $8 \sqrt{6}$ along the $x$, $y$ and $z$ axes respectively,
corresponding to number density $\rho=\sqrt{2}$.
These system dimensions are compatible with a perfect FCC crystal,
the ground state for this system (\S\ref{app:FCC_ground}).
For melting, we heat the system from an initial configuration equilibrated to $T_\mathrm{low}=1.90$
using a thermostat at $T_\mathrm{high}=2.35$,
while for freezing, we cool the system from an initial configuration equilibrated to $T_\mathrm{high}=2.10$
using a thermostat at $T_\mathrm{low}=1.48$.

For the slab simulations in the monodisperse system,
we fix $T_\mathrm{low}=2.0$ and $T_\mathrm{high}=2.1$,
which straddle the melting point.
We preserve the compatibility of the system dimensions with a perfect FCC crystal
by not attempting to eliminate the initial pressure gradient at the slab boundary in slab simulations,
other than by our choice of narrowly-separated $T_\mathrm{low}$ and $T_\mathrm{high}$.
Instead, we double the number of particles and the system size along the $x$-axis,
as compared to the systems used for homogeneous simulations.
This allows for time between the early-time pressure wave dissipating
and the left and right slab fronts colliding at the periodic boundary 
of the system.
For freezing, we define our frozen slab to be the region with $x$-coordinate satisfying $\left| x \right| < 6$.
For melting, our preparation protocol (\S\ref{app:prep}) requires us to specify the size of the frozen non-slab region instead,
leaving the size of the melted slab region to be an emergent property of our preparation protocol
dependent on the size of the interface.
We define the non-slab region to be that with $x$-coordinate satisfying $\left| x \right| \geq 8$.

For the polydisperse case, we simulate $N=20000$ particles and
fix $T_\mathrm{low}=0.06$ and $T_\mathrm{high}=0.14$
for both homogeneous and slab  simulations.
The homogeneous heating simulations have initial lengths $200.00$ and $100.00$
along the $x$ and $y$ axes respectively,
corresponding to number density $\rho=1$.
At this density, $T_\mathrm{MCT} \approx 0.12$~\cite{scallietThirtyMillisecondsLife2022},
such that (accounting for density changes%
~\cite{senguptaDensitytemperatureScalingFragility2013})
$T_\mathrm{low}$ and $T_\mathrm{high}$ straddle $T_\mathrm{MCT}$.
The homogeneous heating simulations follow these systems to steady state at $T_\mathrm{high}$,
corresponding to (average) system lengths $204.70$ and $102.35$ ($\rho \approx 0.955$) along the $x$ and $y$ axes.
These $T_\mathrm{high}$ equilibrium configurations are then used as the initial $T_0 = T_\mathrm{high}$ states
for the homogeneous cooling simulations.

Our choice of $T_\mathrm{low}=0.06$ and $T_\mathrm{high}=0.14$ comes from practical considerations.
We require $T_\mathrm{low} < \TMCT$
to ensure that $T_\mathrm{low}$ is low enough for facilitation to play an important role in the equilibrium dynamics%
~\cite{keysExcitationsAreLocalized2011}
and for amorphous order to set in \cite{kirkpatrickScalingConceptsDynamics1989,bouchaudAdamGibbsKirkpatrickThirumalaiWolynesScenarioViscosity2004}.
The lower we set $T_\mathrm{low}$, the stronger we expect dynamical facilitation to be
and the slower we expect the dynamics to be.
We choose $T_\mathrm{low} = 0.06$ as a compromise between strong facilitation and dynamics fast enough
that we are able to see appreciable structural evolution on the time scale of a week of run time.
While this value of $T_\mathrm{low}$ corresponds to a less stable ultrastable glass than is typically encountered
in the experimental literature~%
\cite{ruiz-ruizRealtimeMicroscopyRelaxation2023, dalalInfluenceSubstrateTemperature2015},
it was found in \cite{herreroFrontPropagationUltrastable2023} that choosing a much cooler $T_\mathrm{low}$
($T_\mathrm{low}=0.035$) only quantitatively changes the behavior of fronts observed during heating.
This suggests that $T_\mathrm{low}=0.06$ is low enough to capture the physics of extremely stable systems.

At number density $\rho \approx 0.955$ the onset temperature is $T_\mathrm{onset} \approx 0.15$%
~\cite{scallietThirtyMillisecondsLife2022}.
We have checked that increasing $T_\mathrm{high}$, even above
$T_\mathrm{onset}$
does not affect our results for final temperature $T_\mathrm{eq}=0.14$.
However, choosing $T_\mathrm{high} = 0.14$, below $T_\mathrm{onset}$,
ensures that we are probing a potential ultrastable-to-poorly-annealed glass transition,
rather than a transition between an ultrastable glass and a high temperature liquid.

For the slab simulations in the polydisperse system,
we define the slab to be the region satisfying $\left| x \right| < b$,
where $b=25.00$ for heating and $25.56$ for cooling.
In slab simulations, we greatly reduce the initial pressure gradient at the slab boundary
by first equilibrating the system in its entirety to $T_\mathrm{eq}$,
with system dimensions $200.00$ and $100.00$ if $T_\mathrm{eq}=T_\mathrm{low}$
or $204.70$ and $102.35$ if $T_\mathrm{eq}=T_\mathrm{high}$,
then shrinking or stretching the non-slab region along the $x$-axis such that
$\rho \approx 0.955$ in this region if $T_0 = T_\mathrm{high}$ or $\rho = 1$ if $T_0 = T_\mathrm{low}$,
before finally evolving this non-slab region with Monte Carlo moves at temperature $T_0$ until it reaches steady state.

Our MD simulations are performed in LAMMPS~\cite{plimptonFastParallelAlgorithms1995} (see \S\ref{app:MD}),
adopting a Nos\'e-Hoover thermostat and a Berendsen barostat
that keeps the relative sizes of the system along each axis constant.
In order to minimize the perturbation to the system due to the start-up of the barostat,
we fix the pressure such that the virial contribution to the pressure at the beginning of the MD simulation
matches that of the target $T_\mathrm{eq}$ equilibrium state.
For any given MD simulation, we average our results over $100$ realizations,
except when explicitly considering individual realizations
or when performing the homogeneous simulations on the polydisperse glass,
in which case we average over only $10$ realizations.

We define time ${t=0}$ as the time at which the MD simulations begin.
For our analysis, it will be necessary to measure the distributions of structural
and dynamical quantities in the initial, untransformed state during homogeneous simulations,
as well as at equilibrium at the target temperature $T_\mathrm{eq}$.
Our methodology for obtaining these distributions can be found in \S\ref{app:t0_tinfty}.

\section{Crystal Case Study \label{sec:crystal}}

We first validate our tests by applying them to the first-order melting
and freezing transitions of a monodisperse system of particles,
in which the growth of nuclei is driven by bulk free energy differences~\cite{pengSituObservationCoalescence2023},
with structural frustration at the boundary of the nucleus suppressing roughness
and promoting domain growth~\cite{pennImperfectOrientedAttachment1998, pennOrientedAttachmentGrowth1998}.
We track the evolution of structure in this system using
Tong and Tanaka's $\Theta$ order parameter~\cite{tongRevealingHiddenStructural2018},
a local measure of packing inefficiency that vanishes in the limit of perfect FCC structure.
We also track the corresponding evolution of the mobility $\mu \left( t \right)$,
defined here for each particle as the magnitude of its inherent state displacement
between times $t$ and $t+0.1$.

\begin{figure*}
\includegraphics[width=\textwidth]{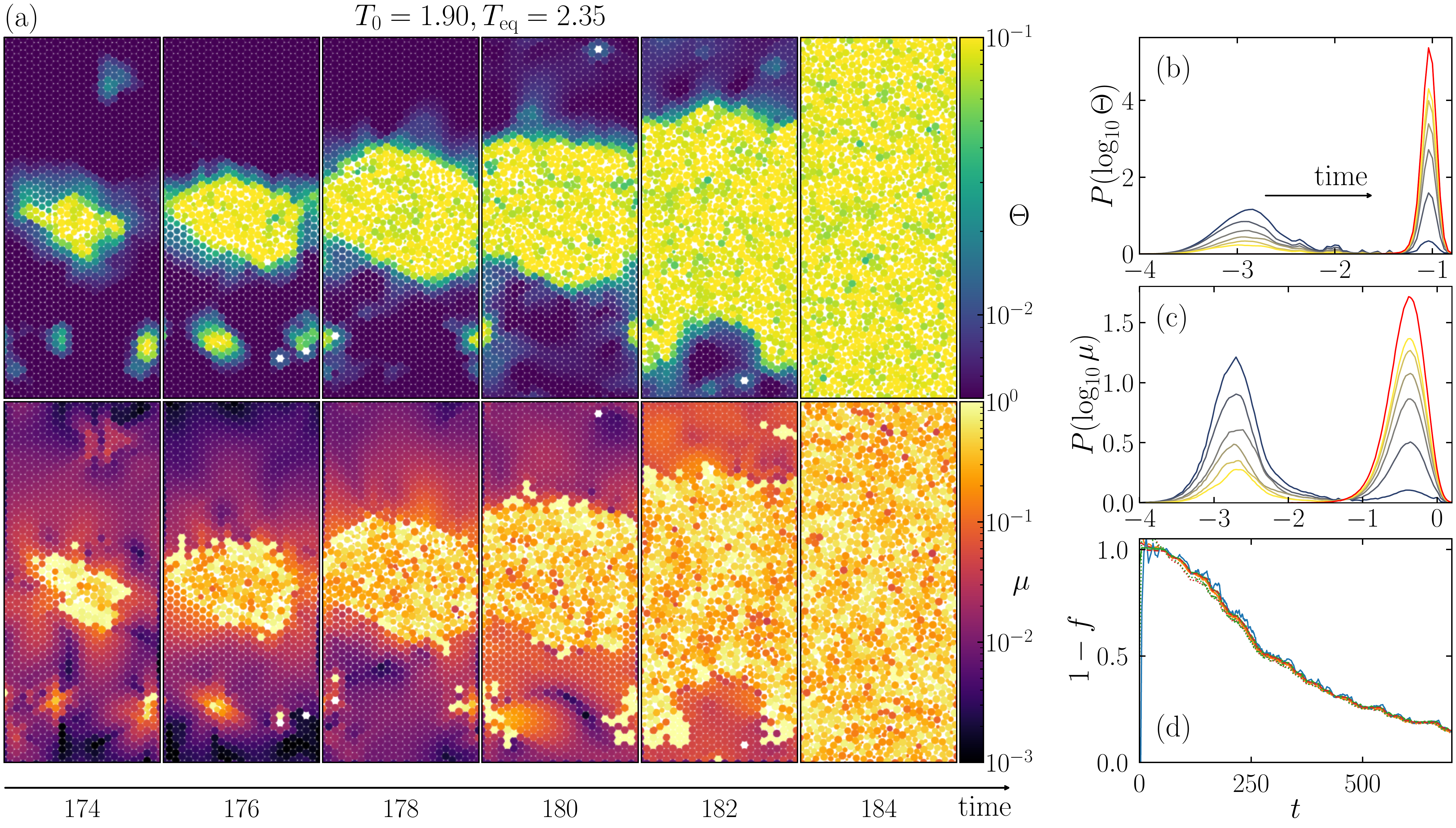} \\
\includegraphics[width=\textwidth]{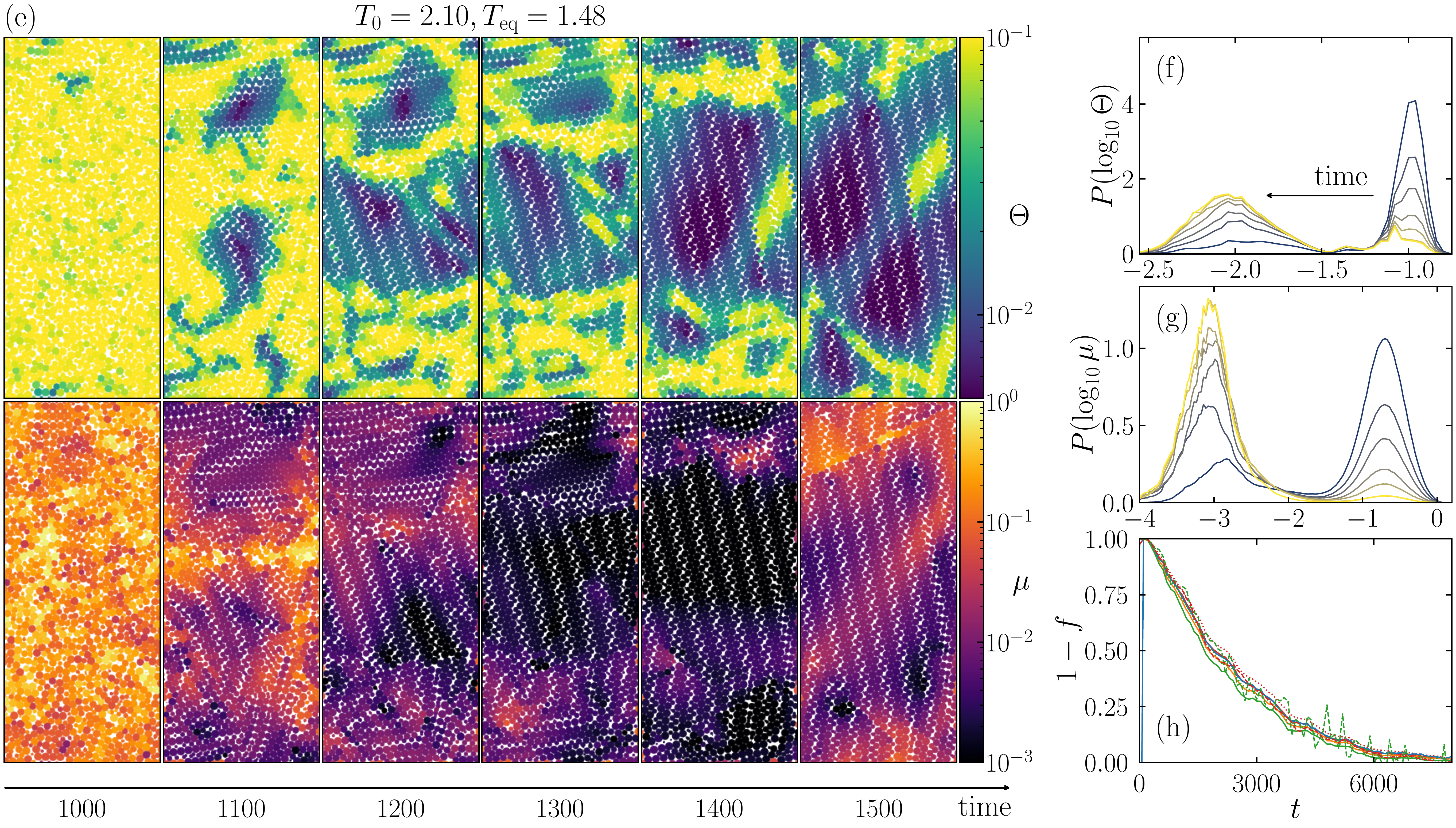}
\caption{Crystal melting (a--d) and freezing (e-h) transitions in a monodisperse system.
         (a, e): snapshots in the {$z \equiv 0$} plane of (top row) the $\Theta$ order parameter
         and (bottom row) the mobility $\mu$ as a nucleus appears and grows
         in an individual trajectory.
         (b, c, f, g):
         distributions (b, f) $P \left( \log_{10} \Theta \right)$
         and (c, g) $P \left( \log_{10} \mu \right)$
         at times (b, c) $t=100$, $200$, $\dots$, $600$ and
         (f, g) $t=1000$, $2000$, $\dots$, $7000$ (later time, lighter color).
         The red curves in (b) and (c) are equilibrium distributions.
         (d, h): untransformed fraction $1-f$ for $f$ calculated from
         $P \left( \log_{10} \Theta \right)$ (dashed red),
         $P \left( \log_{10} \mu \right)$ (dashed green), potential energy
         $\overline V$ (solid orange),
         inherent state potential energy $\overline\VIS$ (solid blue),
         $\overline\Theta$ (solid red),
         $\overline{\log_{10} \Theta}$ (dotted red),
         $\overline{\mu}$ (solid green),
         and $\overline{\log_{10} \mu}$ (dotted green).
\label{fig:homog_crystal}}
\end{figure*}

\emph{Test 1:} Here we simply look for the emergence and growth
of $T_\mathrm{eq}$-structure domains in individual realizations of the homogeneous simulation.
Figs.~\ref{fig:homog_crystal}(a) for melting and \ref{fig:homog_crystal}(e) for freezing
(see also \S\ref{app:supp_movies})
both show a domain of $T_\mathrm{eq}$ structure and dynamics
which grows to span the system along the narrow horizontal axis.
At this point, the domain can expand without gaining any more interfacial area,
making this width an upper bound for the critical nucleus size,
and we see the domain quickly grow to envelop the entire system.

\emph{Test 2:} Here we look for evidence of a waiting time for nucleation.
In Fig.~\ref{fig:homog_crystal}(a), we see that the time between forming a system-spanning nucleus
and melting the entire system is of order $t \sim 10^0$,
while it takes a time of order $t \sim 10^2$ for this system-spanning nucleus to appear.
For cooling (Fig.~\ref{fig:homog_crystal}(e)) these time scales are $t \sim 10^2$ and $t \sim 10^3$, respectively.
This scale separation between the nucleation and domain growth times
holds true across realizations (\S\ref{sec:discussion}).

\emph{Two-state scenario:} Our third and fourth tests involve slab simulations.
To ensure that results from these are relevant to systems without an artificial slab,
we first demonstrate that the structural evolution observed in the homogeneous simulations
corresponds to a two-state scenario, such that at any time $t$,
a fraction $f$ of the system has local structure distributed according to equilibrium at the thermostat temperature $T_\mathrm{eq}$
(the target state),
while the remaining fraction $1-f$ retains its initial distribution of structure,
corresponding to equilibrium at temperature $T_0$ (the initial state).
In this scenario,
widely expected to hold for the heating of ultrastable glasses%
~\cite{kearnsObservationLowHeat2010, rodriguez-tinocoEvaluationGrowthFront2014,
rodriguez-tinocoTransformationKineticsVapordeposited2015, rafols-ribeRoleThermodynamicStability2017,
cubetaCommunicationSurfacefacilitatedSoftening2017, rodriguez-tinocoSurfaceBulkInterplayVaporDeposited2019,
vila-costaNucleationGrowthSupercooled2020,
vila-costa_emergence_2023},
the distribution $P \left( \phi, t \right)$
of a local structural indicator $\phi$ with initial-state distribution $P_0 \left( \phi \right)$
and target-state distribution $P_\mathrm{eq} \left( \phi \right)$ will evolve according to
\begin{equation}
    P \left( \phi, t \right) = \left( 1-f (t) \right) P_0 \left( \phi \right) + f (t) P_\mathrm{eq} \left( \phi \right), %
    \label{eq:avrami}
\end{equation}
interpolating between $P_0 \left( \phi \right)$ and $P_\mathrm{eq} \left( \phi \right)$
as the transformed fraction $f$ grows with time.
Given distributions $P_0 \left( \phi \right)$, $P_\mathrm{eq} \left( \phi \right)$,
and $P \left( \phi, t \right)$ for a structural quantity $\phi$,
we can extract the transformed fraction $f$ from $P \left( \phi, t \right)$
under the assumption that Eq.~\ref{eq:avrami} holds (see \S\ref{app:distbn_interp}).
Alternatively, Eq.~\ref{eq:avrami} implies that
\begin{equation}
    1-f \left( t \right) = \frac{\overline\phi_\mathrm{eq} - \overline \phi \left( t \right)}{\overline\phi_\mathrm{eq} - \overline\phi_0},
    \label{eq:mean}
\end{equation}
where $\overline\phi_0$, $\overline\phi_\mathrm{eq}$, and $\overline \phi \left( t \right)$
are the mean values of the quantity $\phi$ distributed according to $P_0 \left( \phi \right)$,
$P_\mathrm{eq} \left( \phi \right)$, and $P \left( \phi, t \right)$, respectively,
allowing us to calculate $f$ from the evolving mean of $\phi$.

If the two-state scenario holds, we should obtain the same value for the transformed fraction $f$,
regardless of whether we use the distribution $P \left( \phi, t \right)$
or the mean $\overline\phi \left( t \right)$ to calculate it,
and regardless of which local structural variable $\phi$ we choose.
We can therefore test the validity of the two-state scenario by calculating $f$
from both $P \left( \phi, t \right)$ and $\overline\phi$
for a variety of different structural quantities $\phi$
(\S\ref{app:twoStateScenario}).
In practice, we only calculate $f$ from $P \left( \phi, t \right)$ for a single structural quantity $\phi$,
since this is more complicated to do (\S\ref{app:distbn_interp}).
Though it is not a requirement of the two-state equilibration scenario,
we also check if we can obtain $f$ from $P \left( \phi, t \right)$ and $\overline\phi$
when $\phi$ is a dynamical, rather than structural, quantity,
since the coupling of structure and dynamics is a question of physical interest.

In Figs.~\ref{fig:homog_crystal}(b), \ref{fig:homog_crystal}(c),
\ref{fig:homog_crystal}(f), and \ref{fig:homog_crystal}(g),
we see that the $T_0$ and $T_\mathrm{eq}$ distributions of $\log_{10} \Theta$ and $\log_{10} \mu$ are well-separated
~\footnote{
    Note that $P_0 \left (\log_{10} \Theta \right)$ for heating is
    not the same as $P_\mathrm{eq} \left( \log_{10} \Theta \right)$ for cooling,
    since a randomly-nucleated crystal is unlikely to be oriented in a direction compatible with the system axes,
    so equilibrium at the perfect FCC crystal structure is not generally attainable upon cooling.
},
making the validity of the two-state scenario immediately apparent.
We calculate $1-f$ from these distributions and plot the resulting curves against $1-f$ calculated from $\overline\phi$
for $\phi$ corresponding to $\Theta$, $\log_{10} \Theta$, $\mu$, $\log_{10} \mu$,
the potential energy $V$ associated with each particle,
and the inherent state potential energy $V_\mathrm{IS}$ associated with each particle.
As shown in Figs.~\ref{fig:homog_crystal}(d) and \ref{fig:homog_crystal}(h),
these different methods of calculating $1-f$ agree well with one another,
confirming validity of the two-state scenario and therefore the soundness of the slab simulation as a test for domain growth,
and showing that structure and dynamics are coupled together in crystal freezing and melting.

\begin{figure}
\includegraphics[width=\columnwidth]{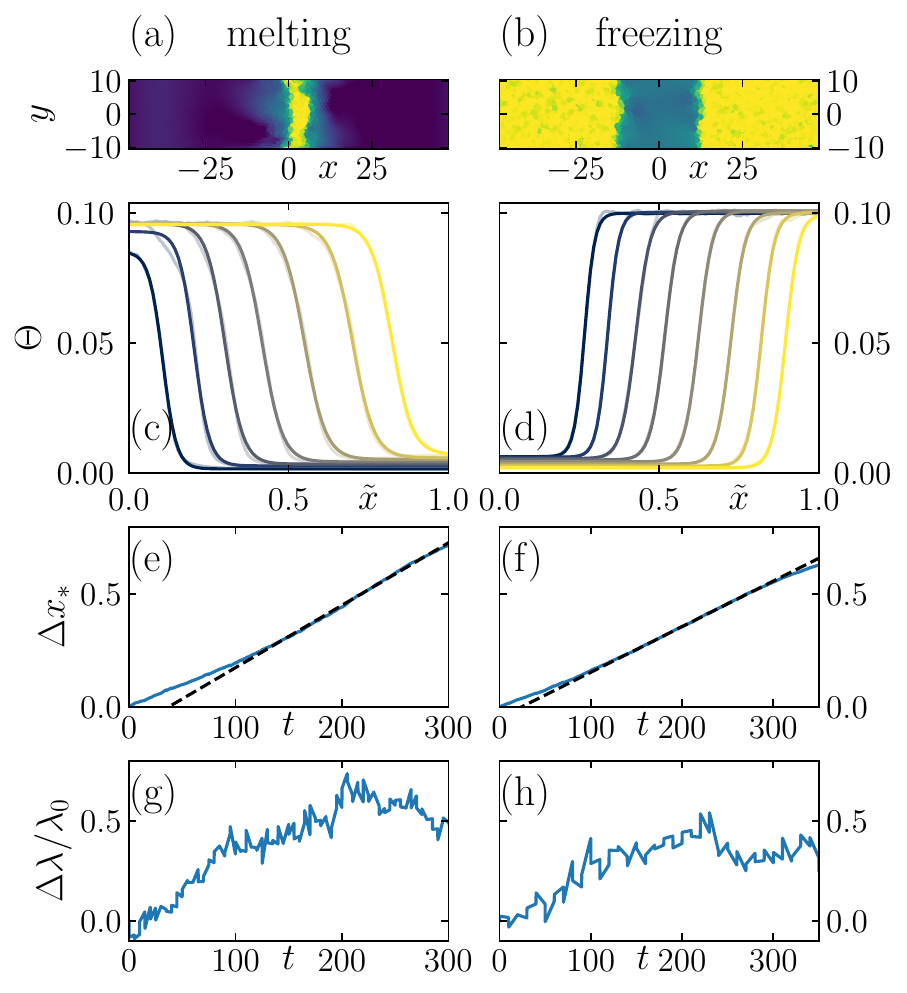}
\caption{Dynamics of melting and freezing for temperatures
         $T_\mathrm{low}=2.0$ and $T_\mathrm{high}=2.1$ tightly straddling the critical point,
         given an artificial target-state slab at time $t=0$.
         (a, b): slice along $z=0$ of the initial configuration,
         colored according to $\Theta$ on a logarithmic scale
         from $10^{-3}$ to $10^{-1}$
         (higher $\Theta$, lighter color).
         (c, d): average $\Theta \left( \tilde x \right)$
         at times $t=0$, $50$, $\dots$, $300$ for melting,
         and $t=0$, $40$, $\dots$, $400$ for freezing
         (later time, lighter color).
         Transparent curves correspond to raw data,
         solid curves to fits.
         Here, $\tilde x = 2 \left| x \right| / L_x$,
         where $L_x$ is the time-dependent system size along the $x$-axis.
         (e--h): change, relative to their initial value,
         of the (e, f) interface center $x_\ast$ and
         (g, h) relative width $\lambda / \lambda_0$
         obtained when fitting the profiles to Eq.~\ref{eq:logistic}.
         Here, the initial interface width $\lambda_0$ has values
         $\lambda_0 = 0.0213$ and $0.0159$ in (e) and (f) respectively. 
         The dashed lines in (e) and (f) are linear fits.
\label{fig:inhom_crystal}}
\end{figure}

Having validated the slab simulations,
we calculate the (ensemble-averaged) $\Theta$ profile along the $x$-axis,
shown as transparent curves in Figs.~\ref{fig:inhom_crystal}(c) and \ref{fig:inhom_crystal}(d).
We identify the interface at the boundary between a $T_\mathrm{eq}$ structure slab and
the surrounding $T_0$ structure region by fitting this profile with a logistic function
\begin{equation}
    \Theta \left( \tilde x \right) = \frac{\Theta_\mathrm{eq} + \Theta_0}{2} %
    + \frac{\Theta_\mathrm{eq} - \Theta_0}{2} \tanh \frac{\tilde x - x_\ast}{2 \lambda},
    \label{eq:logistic}
\end{equation}
where the argument $\tilde x = 2 \left| x \right| / L_x$ accounts for changes in $L_x$
(the system size along $x$) due to the barostat
and takes advantage of the $x \mapsto -x$ symmetry of our slab geometry
\footnote{Symmetry-breaking sample-to-sample fluctuations are suppressed upon ensemble averaging in practice,
and in the limit of infinite size along the $y$ and $z$ axes in principle.}.
The parameters $\Theta_\mathrm{eq}$, $\Theta_0$, $x_\ast$, and $\lambda$ are fit at each snapshot of time $t$,
representing, respectively,
the plateau value of $\Theta$ within the slab of $T_\mathrm{eq}$ structure,
the plateau value of $\Theta$ outside this slab,
the interface center, and the interface width.
The order parameter $\Theta$ is invariant with respect to isotropic expansion,
and is therefore insensitive to the transient effect of the pressure gradient at time $t=0$,
making it an especially suitable choice for our analysis.

\emph{Test 3:} Here we check for domain growth in the slab simulation.
As we see in Figs.~\ref{fig:inhom_crystal}(c) and \ref{fig:inhom_crystal}(d),
the slab of $T_\mathrm{eq}$ structure indeed grows in both melting and freezing simulations.
Figs.~\ref{fig:inhom_crystal}(e) and \ref{fig:inhom_crystal}(f)
show that after an initial transient of around $150$ time units,
which we associate with the dissipation of the pressure gradient present in the initial configuration,
the displacement $\Delta x_\ast = x_\ast \left( t \right) - x_\ast \left( 0 \right)$ of the interface center increases at constant speed.
Both this constant front speed and the two-state equilibration scenario
accord with the description of domain growth within the Avrami formalism%
~\cite{fanfoniJohnsonMehlAvramiKohnogorovModelBrief1998}
for crystallization and melting.

\emph{Test 4}: In our final test we look for evidence that interfacial area is being minimized.
In Fig.~\ref{fig:homog_crystal}(a),
we see a transition from the domain boundary being preferentially aligned along a crystal lattice vector
to being aligned horizontally, while in Fig.~\ref{fig:homog_crystal}(e),
the domain boundary has no particular initial orientation (the surrounding structure is amorphous),
but also quickly becomes horizontal.
This is evidence of an interfacial energy penalty promoting a reduction of interfacial area.
For the slab simulation, the initial width $\lambda_0$ of the interface has to do
with the ability of $T_0$ and $T_\mathrm{eq}$ structure to constrain other structure nearby,
but the two-state equilibration scenario means we can associate
further growth in $\lambda$ with roughening of the interface.
In Figs.~\ref{fig:inhom_crystal}(g) and \ref{fig:inhom_crystal}(h),
after the initial pressure-driven transient ($t \geq 150$),
fluctuations in $\Delta \lambda = \lambda \left( t \right) - \lambda_0$
are comparable in magnitude to the size of the noise floor.
This suppression of roughening is further evidence of an interfacial energy penalty
promoting a reduction of interfacial area.

In summary, the results of our four tests match our expectations for a nucleation and domain growth process,
validating our methodology.

\section{Glasses \label{sec:glasses}}

\begin{figure*}
\includegraphics[width=\textwidth]{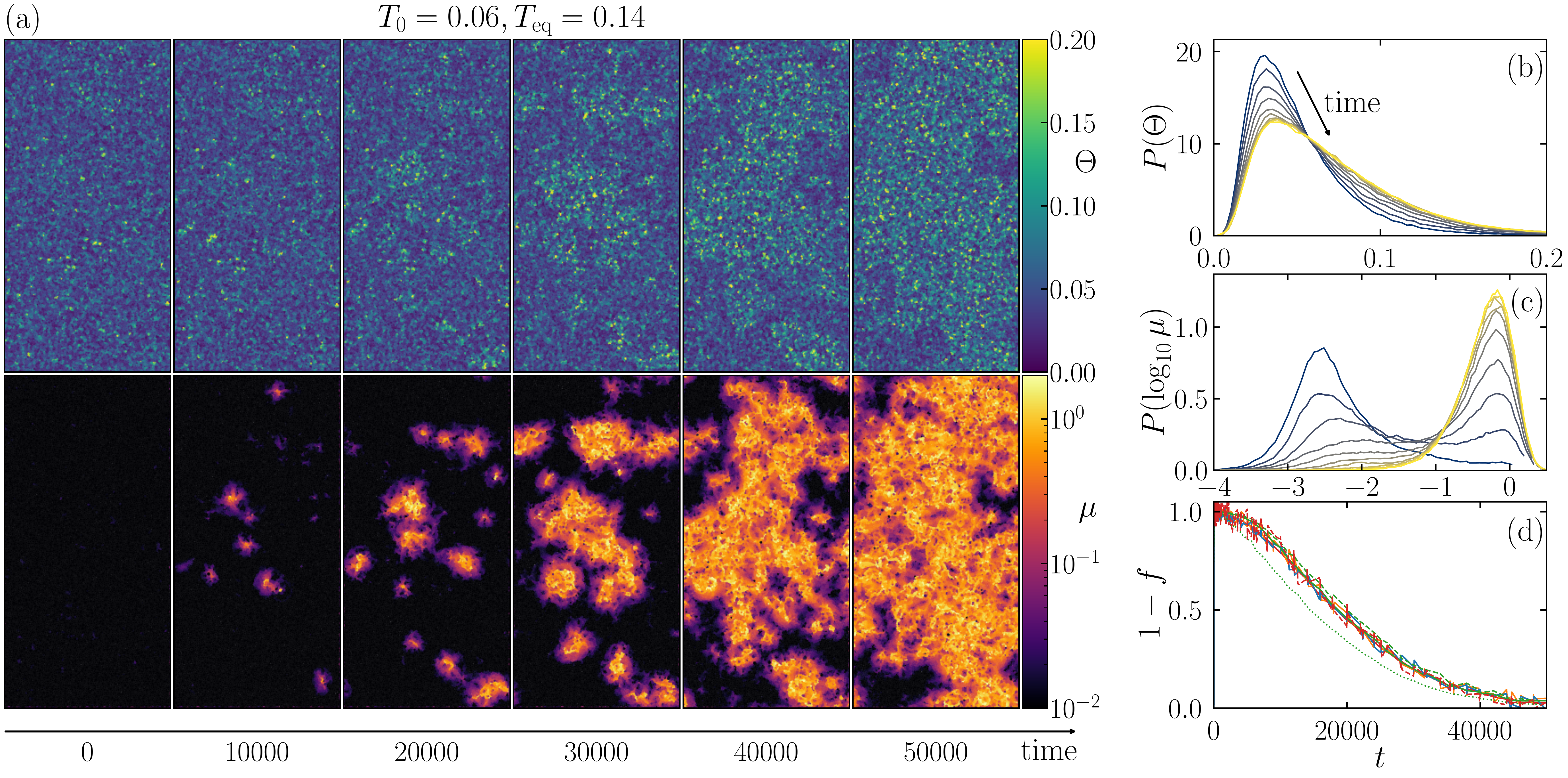} \\
\includegraphics[width=\textwidth]{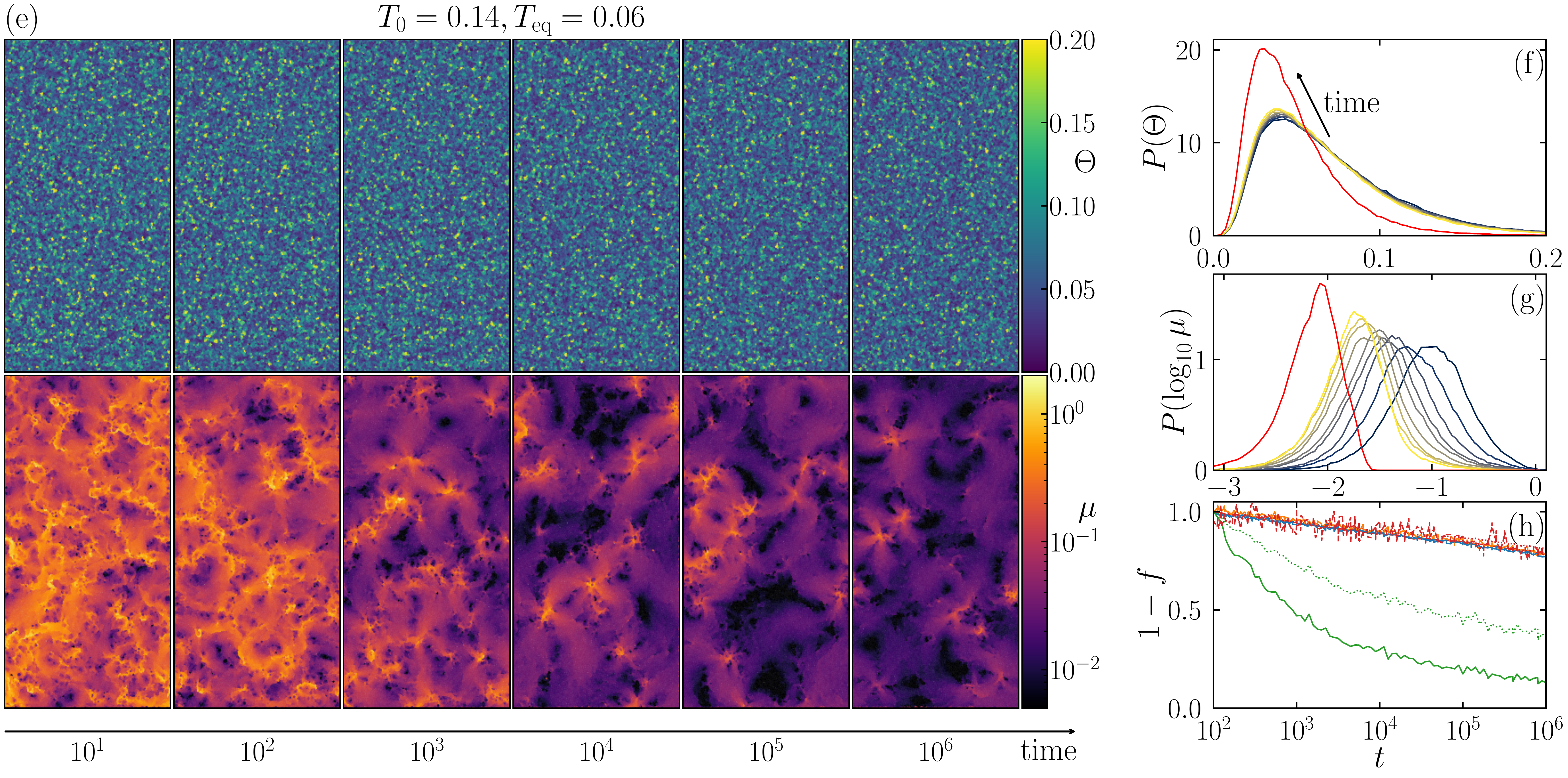}
\caption{Heating (a--d) and cooling (e--h) of a two-dimensional polydisperse supercooled liquid between temperatures
         $0.06$ and $0.14$.
         (a, d): snapshots of the mobility $\mu$ as the system evolves, exhibiting domain growth during heating and
         quadrupoles during cooling, once rearrangements are sufficiently rare.
         (b, c, f, g): evolution of  the (b, f) $\Theta$ and (c, g) $\log_{10} \mu$ distributions at times 
         (b, c) $t=6000$, $12000$, $\dots$, $60000$ and (f, g) $t=10^{2.0}$, $10^{2.5}$, $\dots$, $10^{6.0}$
         (later time, lighter color).
         The red curves in (f) and (g) are equilibrium distributions,
         obtained from equilibrium trajectories at the target temperature and pressure.
         (d, h): untransformed fraction $1-f$ for $f$ calculated from
         $P \left( \log_{10} \Theta \right)$ (dashed red),
         $P \left( \log_{10} \mu \right)$ (dashed green),
         $\overline V$ (solid orange),
         $\overline\VIS$ (solid blue),
         $\overline\Theta$ (solid red),
         $\overline{\log_{10} \Theta}$ (dotted red),
         $\overline{\mu}$ (solid green),
         and $\overline{\log_{10} \mu}$ (dotted green).
\label{fig:homog_poly}}
\end{figure*}

By adjusting our methodology for calculating $\Theta$ (\S\ref{app:Theta})
and redefining the mobility $\mu \left( t \right)$ to be the inherent state displacement of a given particle
between times $t$ and $t + 100$
(accounting for Mermin-Wagner fluctuations as described in \S\ref{app:MW}),
we can use $\Theta$ and $\mu$ to characterize the local structure and dynamics
of our two-dimensional polydisperse glass.

\emph{Test 1:} Here we look for growing $T_\mathrm{eq}$-structure
domains in an individual realization of the homogeneous simulation (\S\ref{app:supp_movies}).
In Fig.~\ref{fig:homog_poly}(a), we see growing high-$\Theta$ and matching high-$\mu$ domains,
representing clear evidence that heating is mediated by domain growth.
We do not, however, see any visible low-$\Theta$ or low-$\mu$ domains during cooling in Fig.~\ref{fig:homog_poly}(e),
suggesting a lack of domain growth in this case.
This is a first and important difference from the crystal case.

\emph{Two-state scenario:} Before applying our second test, which will benefit from the results of the slab simulations,
we confirm the validity of the two-state scenario.
We find good agreement between $1-f$ calculated from the different structural quantities
$P \left( \Theta \right)$, $\overline{V}$, $\overline{\VIS}$, $\overline\Theta$, and $\overline{\log_{10} \Theta}$
in Figs.~\ref{fig:homog_poly}(d) and \ref{fig:homog_poly}(h),
establishing the validity of the two-state scenario for both heating and cooling.
These figures also highlight a major difference between heating and cooling:
the coupling of structure and dynamics.
For heating, this coupling is evident from the matching high-$\Theta$ and high-$\mu$ domains,
and the good agreement between $1-f$ calculated using any of our structural quantities
with $1-f$ calculated from $P \left( \log_{10} \mu \right)$ and $\overline\mu$ (Fig.~\ref{fig:homog_poly}(d))
demonstrates this yet further.
(We do see a deviation from this trend of $1-f$ calculated from $\overline{\log_{10} \mu}$
due to the rightwards shift of the $P_0 \left( \log_{10} \mu \right)$ mode with time in Fig.~\ref{fig:homog_poly}(c);
see also \S\ref{app:perturbed}.)

For cooling, however, $1-f$ calculated from $\overline\mu$ or $\overline{\log_{10} \mu}$
decays much faster than $1-f$ calculated from structural quantities (Fig.~\ref{fig:homog_poly}(h)),
and while $P \left( \Theta \right)$ evolves in accordance with the two-state scenario of Eq.~\ref{eq:avrami}
(Fig.~\ref{fig:homog_poly}(f); see also \S\ref{app:distbn_interp}),
the evolution of $P \left( \log_{10} \mu, t \right)$,
comprising a continuous shift of the initial distribution $P_0 \left( \log_{10} \mu \right)$
towards lower mobility (Fig.~\ref{fig:homog_poly}(g)),
does not.
We therefore see that structure and dynamics are decoupled during cooling.

\emph{Test 2:} Our second test is to look for evidence of a nucleation time.
With no evidence of domain growth during cooling,
we focus here on heating.
In a classical nucleation theory scenario, small domains are suppressed,
so the nucleation time should be much larger than the time scale of individual rearrangements
(successful nucleation should require many attempts),
as was the case for the crystal (Fig.~\ref{fig:homog_crystal}).
We see in Fig.~\ref{fig:homog_poly}(a) that
$T_\mathrm{high}$-structure domains first appear at time $t \lesssim 10^4$.
As can be seen in Fig.~\ref{fig:homog_poly}(d) (see also Fig.~\ref{fig:decays}(c)),
this time scale is no larger than the time scale $t \sim 10^4$
for the relaxation of the system as the domains grow.

\begin{figure}
\includegraphics[width=\columnwidth]{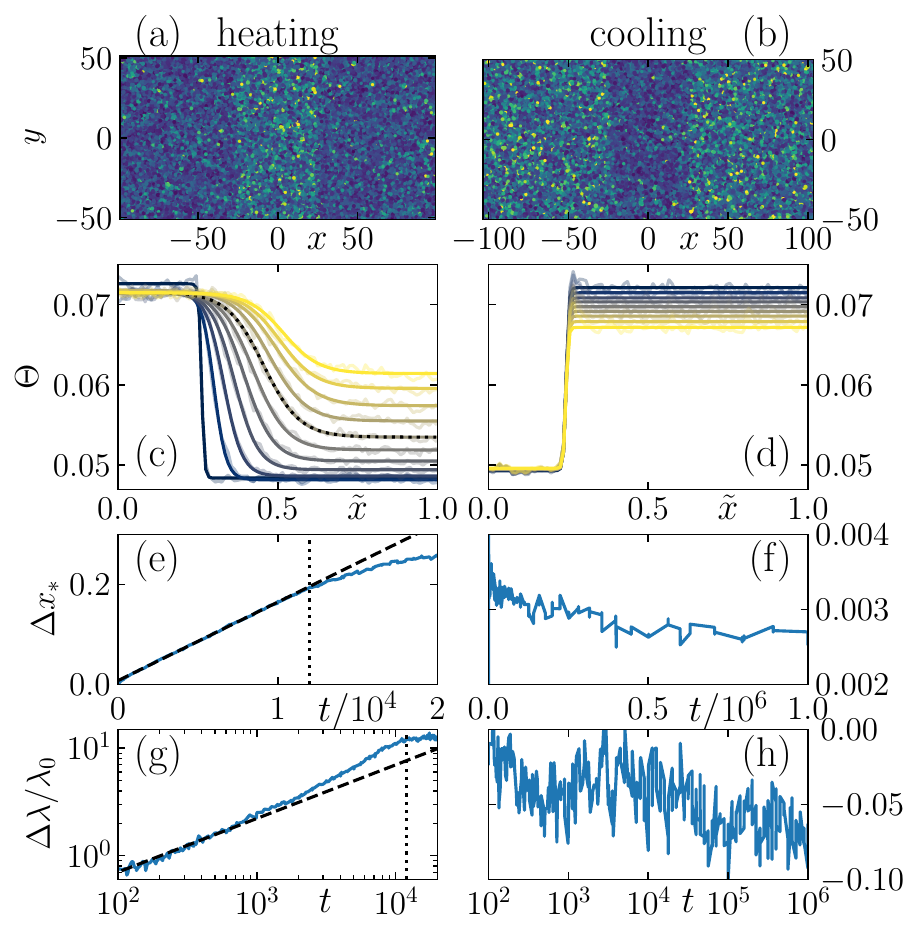}
\caption{Dynamics of (left column) heating and (right column) cooling supercooled liquids
         between temperatures $T_\mathrm{low}=0.06$ and $T_\mathrm{high}=0.14$ straddling $\TMCT$,
         given an artificial target-state slab at time $t=0$.
         (a, b): initial configuration for (a) heating and (b) cooling,
         colored according to $\Theta$ on a linear scale from $0.0$ to $0.1$
         (higher $\Theta$, lighter color).
         (c, d): average $\Theta \left( \tilde{x} \right)$
         at times (c) $t=0$, $2000$, $\dots$, $20000$ and
         (d) $t=10^{2.0}$, $10^{2.5}$, $\dots$, $10^{6.0}$
         (larger time, lighter color).
         Transparent curves correspond to raw data,
         solid curves to fits.
         As before, $\tilde x = 2 \left| x \right| / L_x$,
         where $L_x$ is the time-dependent size of the system along the $x$-axis.
         (e--h): change, relative to their initial value,
         of the (e, f) interface center $x_\ast$ and (g, h)
         relative width $\lambda / \lambda_0$
         obtained when fitting the profiles to Eq.~\ref{eq:logistic}.
         Here, the initial interface width $\lambda_0$ has values
         $\lambda_0 = 0.00428$ and $0.00414$ in (e) and (f) respectively. 
         The dashed line in (e) is a linear fit, while the dashed line in (g)
         represents diffusive growth $\sim t^{\frac{1}{2}}$,
         showing that $\lambda$ is growing super-diffusively.
         The dotted curve in (c), and vertical lines in (e) and (g) correspond to time $t=12000$,
         above which we judge the left and right boundaries of the slab to interact with one another
         at the $\tilde x = \pm 1$ periodic boundary of the system.
\label{fig:inhom_poly}}
\end{figure}

This absence of a significant waiting time (and hence putative nucleation time)
is reinforced by the results of the slab simulations.
In these, we find (Fig.~\ref{fig:inhom_poly}(e)) that $T_\mathrm{high}$-structure fronts
advance with speed $1.57 \times 10^{-3}$ mean particle diameters per time unit into the bulk.
A domain growing isotropically at this speed would take $6.37 \times 10^4$ time units
to span the $x$-axis ($L_x \approx 200$).
This is the time required for a growing, initially point-sized domain
to fully relax the system into the $T_\mathrm{high}$ state,
without allowing for any waiting time.
We see in Fig.~\ref{fig:homog_poly}(d) that this is a good quantitative match
with the actual time taken for the system to fully relax into the $T_\mathrm{high}$ state,
implying that the waiting time is small on the time scale of domain growth.
The high density of disjoint growing domains visible at $t = 20000$
in Fig.~\ref{fig:homog_poly}(a) attests to this too.

\emph{Test 3:} Here we see if $T_\mathrm{eq}$-structure slabs grow in the slab simulation.
As we have already mentioned, ${T_\mathrm{eq}=T_\mathrm{high}}$-structure slabs indeed grow in heating simulations,
with the center of the slab interface, $x_\ast$,
advancing with constant speed up to the system size limit
(Figs.~\ref{fig:inhom_poly}(c) and \ref{fig:inhom_poly}(e)).
In the case of cooling, however, we see that slabs of $T_\mathrm{eq}=T_\mathrm{low}$ structure do not grow at all
(Figs.~\ref{fig:inhom_poly}(d) and \ref{fig:inhom_poly}(f)).
This shows that the reason domains of $T_\mathrm{low}$ structure are not visible in Fig.~\ref{fig:homog_poly}(e)
is that such domains do not grow in the first place.
In a classical nucleation theory scenario, the competition between bulk and interfacial free energies
would apply to both heating and cooling, with the free energy difference between $T_\mathrm{low}$
and $T_\mathrm{high}$ structure in fact being larger during cooling given
the lower thermostat temperature $T_\mathrm{eq} = T_\mathrm{low}$.
This asymmetry and lack of domain growth during cooling is therefore a strong piece of evidence against this scenario.

\emph{Test 4:} Here we investigate whether the size of the interface is being minimized.
The high-mobility domains visible in Fig.~\ref{fig:homog_poly}(a)
clearly have rough edges, and we see no alignment along the horizontal axis
to minimize the length of the interface.
Indeed, spanning the system appears to have no impact on the growth of domains in Fig.~\ref{fig:homog_poly}(a),
unlike in the crystal case,
where spanning the system was associated with an acceleration of domain growth
and alignment of the domain boundaries with the short axis of the system
(Figs.~\ref{fig:homog_crystal}(a) and \ref{fig:homog_crystal}(e)).
The growth of the interface width $\lambda$ in slab simulations provides further evidence
against a thermodynamic drive to minimize the size of the interface.
After an early-time transient ($t \lesssim 10^3$) of diffusive growth ($\Delta \lambda \sim t^{1/2}$),
we see that that the growth in $\lambda$ during heating (Fig.~\ref{fig:inhom_poly}(g))
is super-diffusive until the time $t \approx 12000$
at which the opposite sides of the slab meet at the periodic boundary.
This indicates significant interfacial roughening,
faster even than the na\"ive diffusive expectation for an unsuppressed, freely-diffusing interface.

In summary, the glass behaves very differently from the crystal upon equilibration. All four tests reveal discrepancies between observations and expectations based on a nucleation and growth picture. 

\section{Plaquette Models \label{sec:plaquette}}

Here we show that the very same findings that are inconsistent with a nucleation and growth picture  are explained naturally with dynamical facilitation. We mean dynamical facilitation in its broadest sense as the self-propagation of mobility,
independent of the mechanisms through which this propagation is achieved.
To this end, we apply our four tests to the TPM and SPPM. These models have been shown to display a diverging (at zero temperature) point-to-set length,
but the interfacial tension between amorphous metastable states in these systems is zero and 
their equilibrium dynamics is completely driven by dynamical facilitation,
not by RFOT dynamics~\cite{jackCagingMosaicLength2005}.

\subsection{Model details}

The TPM and SPPM are, respectively, two- and three-dimensional lattice models
comprising spins $\sigma_{\bm{i}} \in \left\{ -1, 1 \right\}$
($\bm{i}$ being the lattice coordinate $\left( i, j \right)$ or $\left( i, j, k \right)$)
interacting via Hamiltonians
\begin{equation}
    H = - \frac{J}{2}\sum_{\bm{i}} p_{\bm{i}},
    \label{eq:Hamiltonian}
\end{equation}
where $J$ is the interaction energy and the plaquette $p_{\bm{i}} \in \left\{ -1, 1 \right\}$
is the product of the triplet of spins $s_{i-1, j}$, $s_{i-1, j+1}$, and $s_{i, j}$
in the TPM and the quintuplet of spins
$s_{i-1, j-1, k+1}$, $s_{i-1, j, k+1}$, $s_{i, j-1, k+1}$, $s_{i, j, k+1}$, and $s_{i, j, k}$
in the SPPM.
We evolve these systems according to Glauber dynamics~\cite{glauberTimeDependentStatistics1963}:
nodes are randomly selected, one at a time,
to undergo a trial flip $\sigma_{\bm{i}} \mapsto -\sigma_{\bm{i}}$
with probability
\begin{equation}
    P \left( \mathrm{flip} \right) = \frac{1}{1 + e^{\beta \Delta E}}
    \label{eq:Glauber}
\end{equation}
of acceptance, where $\Delta E$ is the change in energy that would result from this spin flip
and $\beta = 1 / k_\mathrm{B} T$ for temperature $T$.
Given periodic boundary conditions and a number of particles along at least one axis that is a power of two,
there is a one-to-one correspondence between spins and plaquettes~\cite{garrahanGlassinessConstrainedDynamics2000},
and it is in the plaquette representation that the TPM and SPPM are seen to be KCMs,
with non-interacting Hamiltonians (Eq.~\ref{eq:Hamiltonian}) and
facilitation resulting from the kinetic constraint that trial moves correspond to
a triplet of plaquettes $p_{i,j}$, $p_{i+1, j-1}$, and $p_{i+1, j}$ flipping simultaneously
in the TPM and a quintuplet of simultaneously flipping plaquettes
$p_{i, j, k}$, $p_{i, j, k-1}$, $p_{i, j+1, k-1}$, $p_{i+1, j, k-1}$, and $p_{i+1, j+1, k-1}$
in the SPPM.

We take the indices $i$, $j$, and $k$ to vary along the $x$, $y$, and $z$ axes, respectively,
with bond length $b$ separating adjacent nodes along an axis,
and adopt $b$, $J$, and $J / k_\mathrm{B}$ as our length, energy, and temperature units.
We assume a constant rate of $N$ trial flips per unit of time,
where $N$ is the number of lattice points in the system.
For both the TPM and SPPM, we simulate systems with $128$ lattice points along each axis.

We once again conduct homogeneous and slab simulations.
For the slab simulations, we define our slab to be the set of points with axis coordinate $\alpha$
satisfying $\left| \alpha \right| < 12$, where $\alpha$ is the $x$-coordinate in the TPM case
and the $z$-coordinate in the SPPM case.
Since plaquettes do not interact in these systems (Eq.~\ref{eq:Hamiltonian}),
we equilibrate the slab and non-slab regions by randomly sampling the excitation state of each plaquette site
with equilibrium probability $1 / \left( 1 + e^{1/T} \right)$ of being excited,
where $T=T_\mathrm{eq}$ in the slab and $T_0$ outside the slab.
Because the TPM is not invariant under $x \mapsto -x$ and the SPPM is not invariant under $z \mapsto -z$,
we treat the interfaces on the left ($\alpha < 0$) and right ($\alpha > 0$) sides of the slab separately.

\subsection{Test results}

\begin{figure*}
\includegraphics[width=\textwidth]{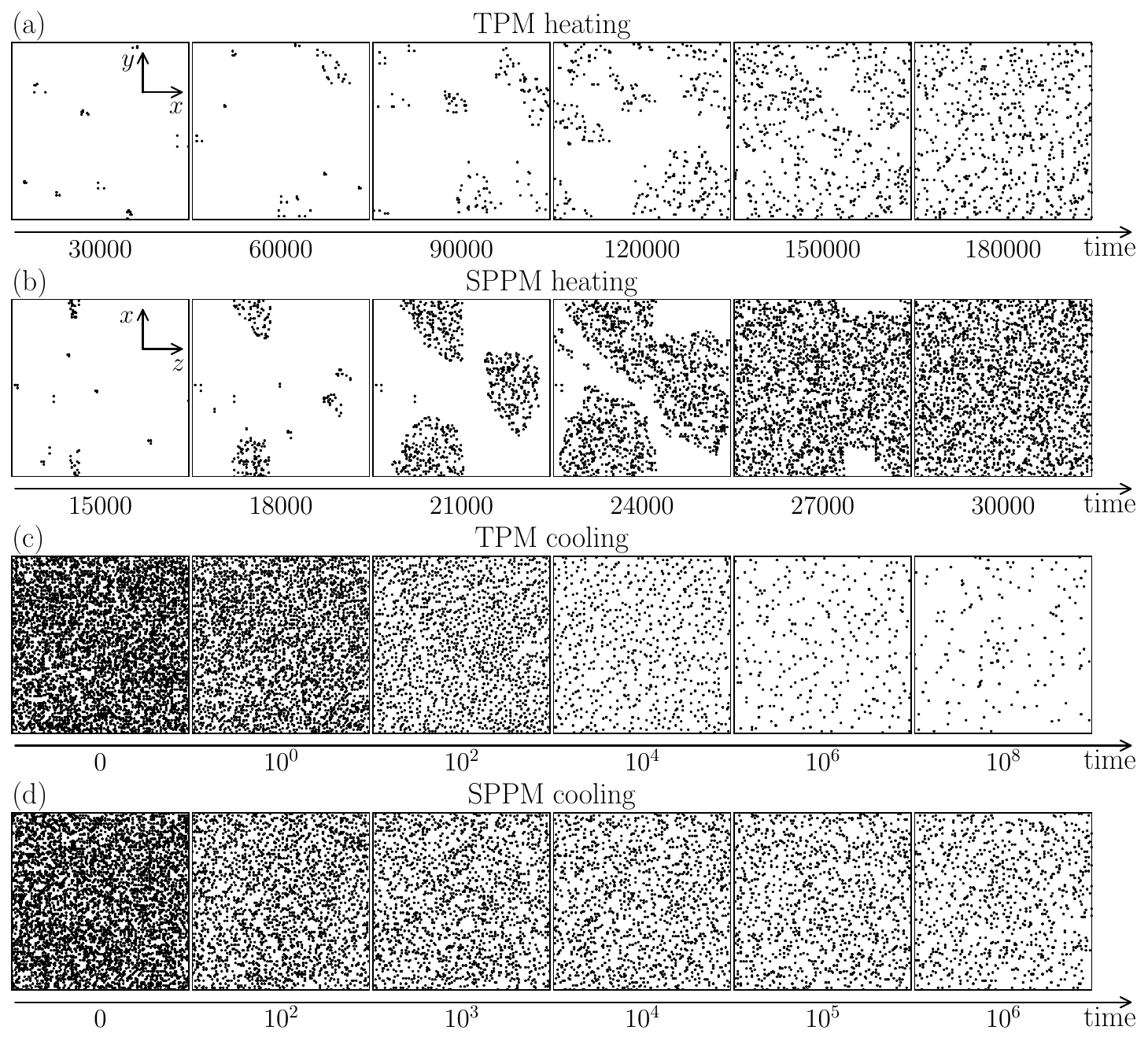} \\
\caption{Structural evolution of the triangular plaquette model (TPM)
         and square pyramidal plaquette model (SPPM) during heating and cooling.
         Systems are initialized at equilibrium at temperature $T_0 = 0.0$
         for heating (a, b) and $1.0$ for cooling (c, d),
         and the system is then evolved with Glauber dynamics at temperature
         $T_\mathrm{eq} = 0.3$ (a), $0.5$ (b), or $0.2$ (c, d).
         Markers correspond to the locations of excited plaquettes
         in the given snapshot of the system.
         In (b), we show a slice along the $y$-plane containing the largest number of excited
         plaquettes at time $t=15000$, so as to capture the birth of the domain.
         In (d), we instead show a slice along the $y$-plane containing the smallest number of
         excited plaquettes at time $t=10^6$, such that if a low-excitation domain
         were present at this time, we would capture it.
\label{fig:homog_plaquette}}
\end{figure*}

\emph{Test 1:} Here we look for growing domains
in an individual realization of the slab simulation.
As we see in Fig.~\ref{fig:homog_plaquette},
there is domain growth in both the TPM and SPPM during heating but not during cooling.
When present, the domains grow with a characteristic triangular shape set by the kinetic constraint,
highlighting the fact that this domain growth is due to dynamical facilitation.
This asymmetry of having domain growth during heating but not during cooling
is consistent with the results of \S\ref{sec:glasses} for the polydisperse glass,
and different from the results of \S\ref{sec:crystal}
for a genuine first-order phase transition in the monodisperse system.

\emph{Test 2:} Here we look for evidence of a nucleation time.
Ironically, given that there is by design no nucleation in these systems
(the plaquettes do not interact, so there can be no interfacial tension),
the high energy barrier to thermal activation in an excitation-free system
means that we do see a significant waiting time of order $t \sim 10^4$
before the emergence of domains, which grow to envelop the system on a time of order $t \sim 10^3$.
This waiting time arises due to the need to escape the large potential energy basin of the defect-free
initial system in order for domain growth to begin.

\begin{figure*}
\includegraphics[width=\textwidth]{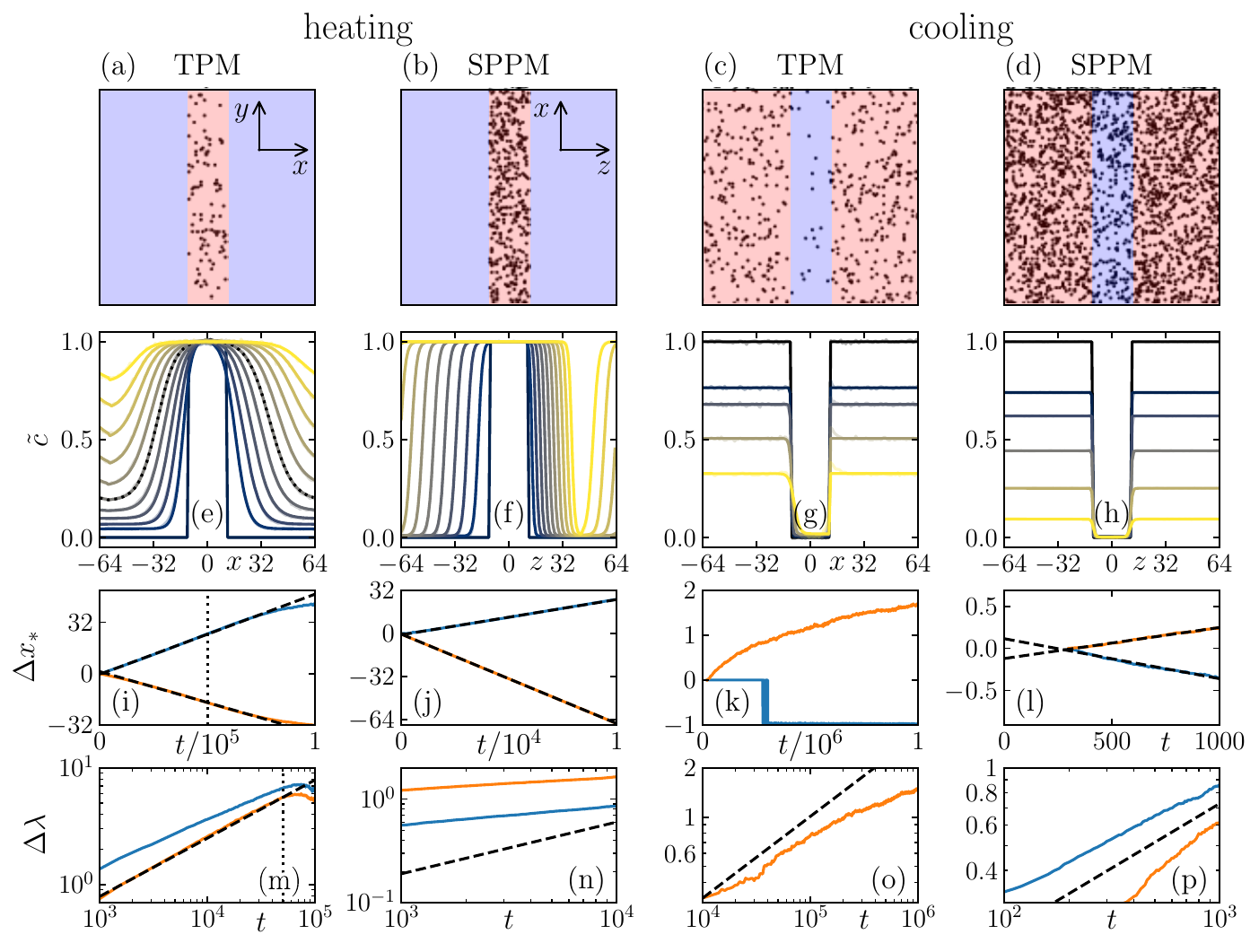} \\
\caption{Dynamics of (left columns) heating and (right columns) cooling
         the TPM and SPPM from temperature $T_0$ to $T_\mathrm{eq}$
         for $\left( T_0, T_\mathrm{eq} \right) = \left( 0.0, 0.3 \right)$ (first column),
         $\left( 0.0, 0.5 \right)$ (second column),
         $\left( 0.3, 0.2 \right)$ (third column),
         and $\left( 0.5, 0.4 \right)$ (last column),
         given an artificial target-state slab at time $t=0$.
         (a--d): initial configuration for
         (a) TPM heating, (b) SPPM heating,
         (c) TPM cooling, and (d) SPPM cooling,
         with markers corresponding to excited plaquettes
         (in the $y=0$ plane for the SPPM).
         The hotter region is shaded red, while the colder region is shaded blue.
         (e--h): normalized concentration
         $\tilde c = \frac{1}{2} + \frac{c - \left( c_\mathrm{eq} + c_0 \right) / 2}{\left| c_\mathrm{eq} - c_0 \right|}$
         of excited plaquettes within bins of constant $x$ (TPM) or $z$ (SPPM)
         at times (e) $t=0$, $10000$, $\dots$, $100000$,
         (f) $t=0$, $1000$, $\dots$, $10000$,
         (g) $t=0$, $10^3$, $10^4$, $10^5$, and $10^6$, and
         (h) $t=0$, $10^1$, $10^2$, $10^3$, $10^4$ and $10^5$
         (larger time, lighter color).
         Here, $c$ is the unnormalized concentration,
         and $c_0$ and $c_\mathrm{eq}$ are the equilibrium concentrations $1 / \left( 1 + e^{1/T} \right)$
         for $T = T_0$ and $T_\mathrm{eq}$, respectively.
         Transparent curves show raw data, solid curves are fits to Eq.~\ref{eq:logistic}.
         The solid curves fully overlap with the transparent curves, apart from in (g).
         Bottom rows: change in the fit parameters (i)--(l) $x_\ast$ and (m)--(p) $\lambda$
         relative to their values in the fit to the $t=0$ profile.
         The orange curves correspond to the slab interface initially at negative $x$,
         the blue curves to the slab interface initially at positive $x$.
         The dashed lines in (i)--(l) are linear fits, the dashed lines in (m)--(p) are power-law fits.
         There is no blue curve in (o) because it is below the noise floor.
         The dotted curve in (e), and vertical lines in (i) and (m) correspond to time $t=50000$,
         above which we judge the left and right boundaries of the slab to interact with one another.
\label{fig:inhom_plaquette}}
\end{figure*}

\emph{Two-state scenario:} The TPM and SPPM comprise non-interacting plaquettes in one of two states,
so we can safely assume that the results of slab simulations will relate to the behavior seen
in the homogeneous simulation.
There is therefore no need to test the two-state scenario as we did for the glass and crystal.

\emph{Test 3:} Here we track domain growth quantitatively in the slab simulation.
As we show in Figs.~\ref{fig:inhom_plaquette}(e), \ref{fig:inhom_plaquette}(f),
\ref{fig:inhom_plaquette}(i), and \ref{fig:inhom_plaquette}(j),
the result of the third test is that slabs in the TPM and SPPM indeed grow upon heating,
with the displacement of the interface center from the slab center increasing at a constant rate.
The speed at which the center of the left interface advances
is different from the corresponding speed for the right interface,
as expected given the asymmetric kinetic constraints.
By contrast, we see in Figs.~\ref{fig:inhom_plaquette}(g), \ref{fig:inhom_plaquette}(h),
\ref{fig:inhom_plaquette}(k), and \ref{fig:inhom_plaquette}(l)
that $T_\mathrm{eq}$-structure slabs in the TPM and SPPM do not grow during cooling,
but sometimes even shrink slightly.
These results are, once again, consistent with the results of the polydisperse glass.

\emph{Test 4:} Here we look for evidence of constraints on the roughness of growing domains.
During heating, the growing TPM domains in Fig.~\ref{fig:homog_plaquette}(a) have rough boundaries,
but we see little roughness in the growing SPPM domains in Fig.~\ref{fig:homog_plaquette}(b).
There is also no evidence of accelerated domain growth once the system spans the box width,
as was seen in Figs.~\ref{fig:homog_crystal}(a) and \ref{fig:homog_crystal}(e) for the monodisperse crystal.
This is confirmed by the growth of the interface width $\lambda$ in the slab simulations.
As shown in Fig.~\ref{fig:inhom_plaquette}(m), $\lambda$ grows approximately diffusively during heating in the TPM,
while, as seen in Fig.~\ref{fig:inhom_plaquette}(n), the growth of $\lambda$ in the heated SPPM is strongly sub-diffusive.
Despite the similar designs of the TPM and SPPM,
the roughness of growing domains during heating has very different dynamics in the two cases.
We note that even within a given model, the power law for the growth of $\lambda$ for the left interface differs
from the power law for the right interface.
This shows that the asymmetric constraint is affecting the rate of roughening,
and highlights the sensitivity of roughening to the mechanism underlying facilitation.
The fact that neither the TPM nor the SPPM see the super-diffusive growth of $\lambda$ seen in the heated polydisperse glass
is consistent with this.

We see that the main phenomena associated with the equilibration dynamics of glasses%
---Avrami-like domain growth during heating and an absence of domains during cooling---%
can be obtained purely via dynamical facilitation, even with trivial thermodynamics,
as in the KCMs studied here,
suggesting that these two-state scenario behaviors are generic to dynamically-facilitated systems.
The one phenomenon seen in the glass but not recovered in the KCMs is super-diffusive roughening during heating,
but this roughening behavior is highly sensitive to the specific details of the system,
varying not only between KCMs, but even between the left and right boundaries of a single slab.

\subsection{Phenomenological model}

From the shared phenomenology of the polydisperse glass and the KCMs,
we can build a minimal, phenomenological model of slab growth,
\begin{equation}
\partial_t c = \left( c_\mathrm{eq} - c \right) / \tau
+ s \left| \partial_x c \right|
+ D \partial_x^2 c,
\label{eq:PDE}
\end{equation}
where $c$ is the excitation state of the system (\emph{e.g.} $\Theta$ or the plaquette concentration $c$),
$\tau$ is the local relaxation time of the system,
$s$ is the speed of mobility propagation (the absolute value ensuring that transport is always directed from high to low mobility),
and $D$ is a diffusion constant accounting for the roughening of the system.
This allows us to discuss the observed dynamics of these systems in terms of specific dynamical ingredients.
For simplicity, we focus on slab growth in the TPM,
identifying the contributions to the rate of change $\partial_t c$ of the plaquette concentration $c$,
and noting system-dependent considerations when they arise.

The first contribution we consider is the mobility propagation $s \left| \partial_x c \right|$.
In the ultrastable glass heating literature, experiments%
~\cite{rodriguez-tinocoTransformationKineticsVapordeposited2015, rafols-ribeRoleThermodynamicStability2017}
and a KCM study~\cite{leonardMacroscopicFacilitationGlassy2010}
find front speeds $s = l_s / \tau_\mathrm{eq} \left( T_\mathrm{eq} \right)^\gamma$,
where $\gamma \approx 1$.
Comparing Figs.~\ref{fig:inhom_poly}(c) and \ref{fig:inhom_plaquette}(f)
with Figs.~\ref{fig:inhom_poly}(i) and \ref{fig:inhom_plaquette}(j),
we see that in both the polydisperse glass and the TPM,
the velocity of the interface center remains constant even after significant relaxation of the non-slab region
towards equilibrium,
so we take $l_s$ to be a constant.

The specific form of the relaxation time $\tau$ is system-dependent,
but any sufficiently fragile system would yield the speeds $s \approx 0$ for low $T_\mathrm{eq}$
and $s > 0$ for high $T_\mathrm{eq}$
seen in Figs.~\ref{fig:inhom_poly} and \ref{fig:inhom_plaquette}.
From the equilibrium behavior $\tau \sim e^{1 / \log (3) T^2}$~\cite{jackCagingMosaicLength2005} of the TPM,
we obtain
\begin{equation}
    \tau = \tau_0 \exp \left( \frac{1 / \log 3}{\left( \left( 1-a \right) T_\mathrm{eq} + a \Tf \right)^2} \right)
    \label{eq:tau}
\end{equation}
by modifying the Tool, Narayanaswami, and Moynihan (TNM) model~\cite{moynihanDependenceFictiveTemperature1976}
to yield the TPM relaxation in equilibrium and to work in the limit of vanishing fictive temperature
$\Tf = 1 / \log \left( c^{-1} - 1 \right)$.
Here, $\tau_0$ is a constant time scale and $a$ is a fit parameter (see \S\ref{app:TPM_fit}).

\begin{figure}
\includegraphics[width=\columnwidth]{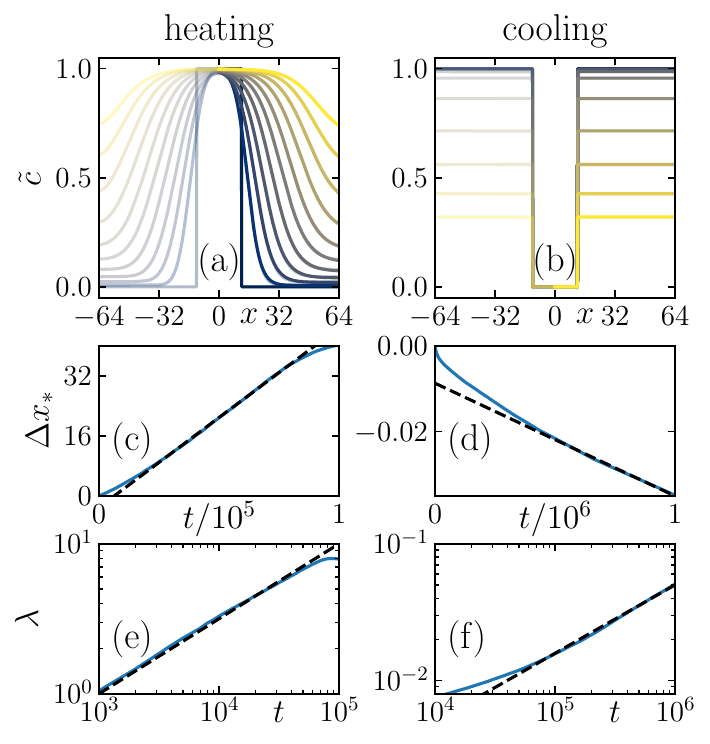}
\caption{Dynamics of the phenomenological model governed by Eq.~\ref{eq:PDE}
         for constants $l_s = 0.18$ and $l_D = 0.25$
         given an artificial target-state slab at time $t=0$.
         Heating (left column) is performed between temperatures
         $T_0 = 0.0$ and $0.3$ using $\tau_0 = 0.549$ and $a = 0.253$
         and cooling (right column) is performed between temperatures
         $T_0 = 0.3$ and $0.2$ using $\tau_0 = 0.705$ and $a = 1.00$.
         Parameters are chosen so as to yield behavior similar to the TPM,
         establishing the reasonableness of this model,
         but no explicit fit to the TPM is performed.
         (a, b): normalized excitation state $\tilde c$
         defined as in Fig.~\ref{fig:inhom_plaquette}(e)
         at times (a) $t=0$, $1000$, $\dots$, $10000$ and
         (b) $t=0$, $10^3$, $10^4$, $10^5$, and $10^6$
         (larger time, lighter color).
         Transparent curves correspond to numerically-solved Eq.~\ref{eq:PDE},
         solid curves to fits to the $x>=0$ part of this solution,
         noting that $c \left( -x \right) = c \left( x \right)$.
         (c--f): change, relative to its initial value,
         of the interface center position $x_\ast$ (c, d),
         and width $\lambda$ (e, f)
         obtained when fitting the profiles to Eq.~\ref{eq:logistic}.
         The dashed lines in (c) and (d) are linear fits,
         while those dashed line in (e) and (f)
         represent diffusive growth $\sim t^{\frac{1}{2}}$.
\label{fig:inhom_model}}
\end{figure}

Finally, the rate of roughening is system-dependent, as we have discussed.
In the TPM, this roughening yields an approximately diffusive growth of the slab width $\lambda \sim t^{1/2}$,
which can be obtained using a diffusive term $D \partial_x^2 c$.
From Figs.~\ref{fig:inhom_poly} and \ref{fig:inhom_plaquette},
we see that when the interface speed is small, so is the diffusion rate,
so we assume proportionality between the two, $D = l_D^2 / \tau_\mathrm{eq}$,
where $l_D$ is a constant.

Comparing Figs.~\ref{fig:inhom_poly} and \ref{fig:inhom_model},
we see that this phenomenological model of relaxation dynamics based on dynamic facilitation captures the key features of slab evolution,
both during heating and during cooling,
justifying the separation of dynamics into distinct local relaxation,
constant-speed drive, and roughening pieces.

\section{Discussion \label{sec:discussion}}

\begin{figure}
\includegraphics[width=\columnwidth]{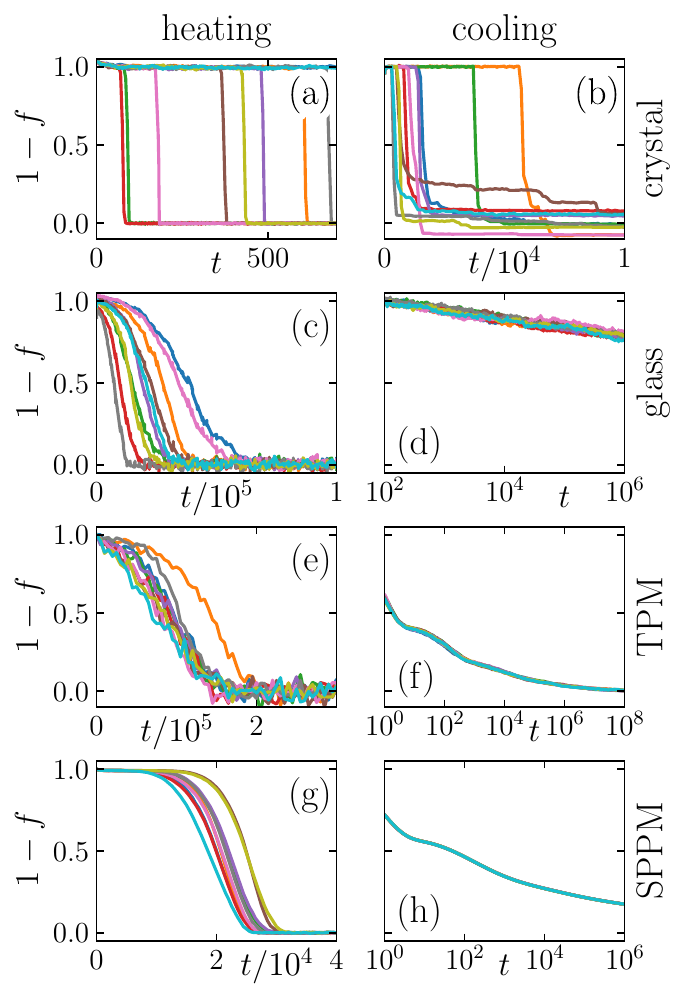}
\caption{(a--d): Evolution of $1-f$ in homogeneous simulations calculated
from Eq.~\ref{eq:mean} using the mean value of $\Theta$ 
for ten individual realizations contributing to the averaged results shown in
(a) Fig.~\ref{fig:homog_crystal}(d), (b) Fig.~\ref{fig:homog_crystal}(h),
(c) Fig.~\ref{fig:homog_poly}(d), and (d) Fig.~\ref{fig:homog_poly}(h).
(e--h): Evolution of $1-f$ calculated from Eq.~\ref{eq:mean} using the concentration $c$ of excited plaquettes
for ten individual realizations in the same regimes as those shown in Fig.~\ref{fig:homog_plaquette}.
\label{fig:decays}}
\end{figure}

We have investigated whether the relaxation of a glass towards equilibrium during heating and cooling
proceeds via nucleation and domain growth in the sense of classical nucleation theory.
By applying the same four tests to the heating and cooling of a three-dimensional monodisperse crystal,
a two-dimensional polydisperse glass, and the TPM and SPPM,
we find that the equilibration dynamics of our glass does not resemble that of the melting or freezing crystal,
but instead resembles that of the dynamically-facilitated plaquette models in every discriminating aspect.

This is made immediately apparent in Fig.~\ref{fig:decays},
which shows the evolution of $1-f$ calculated from Eq.~\ref{eq:mean}
applied to $\Theta$ (\ref{fig:decays}(a)--\ref{fig:decays}(d)) or
the plaquette concentration $c$ (\ref{fig:decays}(e)--\ref{fig:decays}(h))
for ten individual homogeneous simulation trajectories
(\emph{c.f.} the ensemble-averaged data in Figs.~\ref{fig:homog_crystal}, \ref{fig:homog_poly}, and \ref{fig:homog_plaquette}).
For crystal melting and freezing in our monodisperse system, we see a nearly discontinuous
change of $1-f$ in Figs.~\ref{fig:decays}(a) and \ref{fig:decays}(b) once a nucleus is formed,
though freezing trajectories are seen to get trapped in a number of metastable states in Fig.~\ref{fig:decays}(b).
By contrast, the evolution of $1-f$ in both our polydisperse glass and the two plaquette models
(Figs.~\ref{fig:decays}(c)--\ref{fig:decays}(h)) is much smoother and more gradual,
reflecting the absence of a nucleation time in these systems.
For cooling (Figs.~\ref{fig:decays}(d), \ref{fig:decays}(f), and \ref{fig:decays}(h)),
the lack of variation between individual trajectories reflects the lack of domain growth in these systems,
which results in better averaging within individual realizations.

Fig.~\ref{fig:decays} highlights the fact that while the absence of a nucleation time
and the asymmetry between heating being driven by domain growth and cooling being a homogeneous process
is not consistent with a classical nucleation theory scenario,
the equilibration dynamics of our glass is in fact generic to dynamically-facilitated systems.
Individual trajectories of the glass resemble those of the plaquette models and not the crystal because
dynamical facilitation, not nucleation and domain growth, governs the equilibration dynamics of glasses.

\subsection{Nucleation and Domain growth}

While we do find that the equilibration of our glass during heating is mediated by domain growth
(Figs.~\ref{fig:homog_poly}, \ref{fig:inhom_poly}, and \ref{fig:decays}),
we do not find evidence of nucleation for either heating or cooling.
We find no evidence of a nucleation time for the emergence of $T_\mathrm{high}$ domains during heating,
which appear on a similar time scale to the time scale over which the front of a growing $T_\mathrm{high}$ domain
advances by one mean particle diameter.
We also find no evidence of any thermodynamic drive to minimize
the total interface between the domains of $T_\mathrm{low}$ structure and $T_\mathrm{high}$ structure
during this heating process.
Unlike in the crystal case (Figs.~\ref{fig:homog_crystal}(a) and \ref{fig:homog_crystal}(e)),
there is no alignment of the growing $T_\mathrm{high}$ domain in Fig.~\ref{fig:homog_poly}(a)
along the horizontal axis to minimize its size,
and in slab simulations, the slab boundary exhibits super-diffusive roughening.
Cooling ultimately provides the most striking evidence against a drive to minimize the interface between
domains of $T_\mathrm{high}$ and $T_\mathrm{low}$ structure,
since the absence of domain growth combined with the validity of the two-state equilibration scenario
implies that small domains of $T_\mathrm{low}$ structure surrounded by $T_\mathrm{high}$ structure
proliferate during the cooling process.

This result concerns only nucleation in the thermodynamic sense of classical nucleation theory.
First-order transitions in systems coupled to a reference configuration~\cite{berthierEvidenceDisorderedCritical2015}
and dynamical phase transitions~\cite{hedgesDynamicOrderDisorderAtomistic2009}
are found in KCMs,
and an explanation of relaxation dynamics based on those transitions may be possible%
~\cite{turnerOverlapActivityGlass2015, jackMeltingStableGlasses2016}.
As we have shown, however, this explanation cannot involve traditional nucleation and growth,
and we believe that it would ultimately boil down to a complementary view on how dynamical facilitation influences relaxation dynamics.
Note, moreover, that our results also do not contradict the existence of a finite point-to-set length $\xiPTS$,
nor the possibility of an RFOT mosaic.
In fact, the asymmetry between heating and cooling, including the decoupling of structure and mobility during the latter process,
can be derived via the microscopic RFOT theory \cite{lubchenkoTheoryStructuralGlasses2007, wolynesSpatiotemporalStructuresAging2009}.
As shown in \S\ref{app:struct_perturb}, individual rearrangements perturb the surrounding structure
over a length scale $\xi_\mathrm{perturb} \sim 10^0$
similar to the point-to-set length $\xiPTS \sim 10^0$ for our temperature range~\cite{berthierZerotemperatureGlassTransition2019},
with the typical size of perturbations to $\Theta$ comparable to the size of $\Theta$ itself.
This raises the possibility that individual rearrangements create individual domains of $T_\mathrm{low}$ structure
within a mosaic as the system cools.

\subsection{Dynamical facilitation}

We have established that dynamical facilitation drives the domain growth
observed when heating the polydisperse glass.
Despite very different mechanisms for dynamical facilitation,
the interface center $x_\ast$ evolves in the same manner in both the polydisperse glass and our two plaquette models.
In all cases its behavior is  well-captured by a constant-speed driving term
$s \left| \partial_x c \right|$ in Eq.~\ref{eq:PDE}.

In recent work, Herrero, Ediger and Berthier~\cite{herreroFrontPropagationUltrastable2023}
conducted very similar slab simulations to ours using the same polydisperse system and find,
suggestively, that it is individual rearrangements near the slab boundary that advance the transformation front.
In light of the two-state equilibration scenario,
this explains why the significant relaxation of the non-slab region at times $t \geq 10^3$
in the slab heating simulations (Fig.~\ref{fig:inhom_poly}(e))
does not affect the constant front speed.
As this region relaxes, domains of $T_\mathrm{high}$ structure grow
which contribute to raising the height of the $T_\mathrm{low}$-region
plateau in Fig.~\ref{fig:inhom_poly}(c) (parameter $\Theta_0$ in Eq.~\ref{eq:logistic}),
but which do not advance the front.
Instead, the front advances via the transformation of $T_\mathrm{low}$-structure adjacent to the front
facilitated by rearrangements near the interface.
The same scenario explains observations of the same behavior in the TPM (Fig.~\ref{fig:inhom_plaquette}(i)).

The interface width $\lambda$, for its part, provides system-specific information.
The initial interface widths $\lambda_0$ relate to the ability
of local structure to constrain structure nearby,
hence $\lambda_0$ in the crystal (Fig.~\ref{fig:inhom_crystal})
being larger than in the glass (Fig.~\ref{fig:inhom_poly}),
while $\lambda_0 \approx 0$ in the non-interacting plaquette systems
(Fig.~\ref{fig:inhom_poly}).
Given the two-state equilibration scenario, the evolution of roughness
is captured by the evolution of $\lambda$ in the ensemble- and spatially-averaged
profile $\Theta \left( \tilde x \right)$ (Eq.~\ref{eq:logistic}) in slab simulations.
As we have seen, this is strongly system-dependent,
to the point where the widths of the interfaces on the left and right sides of TPM and SPPM slabs
grow as power laws $\lambda \sim t^\nu$ with different power law exponents $\nu$.
Among our glass and two plaquette models,
we find exponents $\nu$ spanning values well above (Fig.~\ref{fig:homog_poly}(g))
to well below (Fig.~\ref{fig:homog_plaquette}(n))
the na\"ive diffusive exponent $\nu = 1/2$ for roughness arising from
random fluctuations in the local front speed.

It was shown in \cite{herreroFrontPropagationUltrastable2023}
that slab roughness in the slab-simulation heating of very cold systems at very long times
reaches a steady state with characteristic length scale set by the dynamic correlation length
$\xi_4$~\cite{lacevicSpatiallyHeterogeneousDynamics2003}.
Future work may similarly uncover the physical origin of the power-law exponent $\nu$.

\subsection{Technical advances}

Our investigations of nucleation and domain growth and dynamical facilitation were
aided by two key technical advances with implications of their own.
The first was the use of Eqs.~\ref{eq:avrami} and \ref{eq:mean}
to confirm the two-state equilibration scenario.
While, on the basis of heat capacity measurements, this
two-state scenario has long been expected to hold for the heating of ultrastable glasses%
~\cite{kearnsObservationLowHeat2010, rodriguez-tinocoEvaluationGrowthFront2014,
rodriguez-tinocoTransformationKineticsVapordeposited2015, rafols-ribeRoleThermodynamicStability2017,
cubetaCommunicationSurfacefacilitatedSoftening2017, rodriguez-tinocoSurfaceBulkInterplayVaporDeposited2019,
vila-costaNucleationGrowthSupercooled2020,
vila-costa_emergence_2023},
our work here contains the most direct confirmation of the two-state scenario to date. 

Remarkably, we find that the two-state scenario holds for
\emph{cooling} as well as heating, despite the lack of domain growth in the former.
This has encouraging implications for the modelling of glasses out of equilibrium,
showing that local structure after a rearrangement can simply be sampled directly and independently
from the equilibrium distribution of structure at the target equilibrium temperature,
as for instance assumed in trap models~\cite{monthusModelsTrapsGlass1996}.

A second technical advance of our work is the introduction of the phenomenological model of slab growth, Eq.~\ref{eq:PDE}.
This model shows how slab growth during heating can be reconciled with homogeneous relaxation during cooling
given the same set of dynamical ingredients.
By identifying the distinct components of the dynamics of equilibrating glasses,
this model should also aid the development of future models for the equilibration of glasses in non-slab geometries.

\section*{Acknowledgements}
The authors thank C. Scalliet for sharing data sets
from \cite{scallietThirtyMillisecondsLife2022} with them.
R.N.C. is grateful to L. Berthier, H.-H. Boltz, R.C. Dennis, Z. Fakhraai,
M.A. Galvani Cunha, C. Herrero, P. Luo, P.D. Olmsted, M. Ozawa, and A.G. Yodh
for helpful discussions concerning this project.
The design of Figs.~\ref{fig:homog_crystal} and \ref{fig:homog_poly} in this paper was
heavily inspired by the figures in \cite{chardacTopologyDrivenOrderingFlocking2021}.
This work was supported by the Simons Foundation via the ``Cracking the glass problem'' collaboration (\#454945, RNC and AJL).
AJL also thanks the Simons Foundation for support via \#327939
as well as for the hospitality of the Center for Computational Biology at the Flatiron Institute,
as well as the Isaac Newton Institute for Mathematical Sciences at Cambridge University (EPSRC grant EP/R014601/1),
for support and hospitality.
This work used the Anvil Supercomputer at the Rosen Center for Advanced Computing through allocation PHY200101
from the Extreme Science and Engineering Discovery Environment (XSEDE)~\cite{XSEDE},
which was supported by National Science Foundation grant number \#1548562.

\appendix

\section{Methodological Details}

\subsection{Swap potential \label{app:swap}}

For convenience, we note here that in Eq.~\ref{eq:pair_potential},
\begin{align}
    c_0 &= - 28 \tilde r_\mathrm{cut}^{-12} \\
    c_1 &= 48 \tilde r_\mathrm{cut}^{-14} \\
    c_2 &= -21 \tilde r_\mathrm{cut}^{-16},
\end{align}
where $\tilde r_\mathrm{cut} = 1.25$.
We also note that $A$, $\sigma_\mathrm{min}$, and $\sigma_\mathrm{max}$ in the
probability density function
\begin{equation}
    P \left( \sigma \right) = 
    \begin{cases}
        A \sigma^{-3} & \sigma \in \left[ \sigma_\mathrm{min}, \sigma_\mathrm{max} \right]  \\
        0 & \mathrm{otherwise}
    \end{cases}
\end{equation}
for the particle size $\sigma$ can be found by solving
\begin{equation}
    2 A \tanh \left( \frac{\overline\sigma^2 \left( 1 + c_\sigma^2 \right)}{2 A} \right) = \overline\sigma^2
\end{equation}
numerically for $A$ and noting that
\begin{equation}
    \frac{\sigma_\mathrm{min}}{\sigma_\mathrm{max}} = \exp \left( -\frac{\overline\sigma^2 \left( 1 + c_\sigma^2 \right)}{A} \right)
\end{equation}
and
\begin{equation}
    \overline\sigma = A \left( \sigma_\mathrm{min}^{-1} - \sigma_\mathrm{max}^{-1}  \right),
\end{equation}
ultimately yielding
\begin{align}
    A &\approx 1.3145271918 \\
    \sigma_\mathrm{min} &\approx 0.7244461244 \\
    \sigma_\mathrm{max} &\approx 1.6138530488
\end{align}
for $\overline\sigma = 1$ and $c_\sigma = 0.23$.

\subsection{Preparation protocols \label{app:prep}}

To homogeneously equilibrate to low temperatures,
we conduct Monte Carlo simulations at constant volume and temperature until steady state,
randomly selecting particles and conducting trial displacements with components
randomly sampled from the interval $\left( -0.05, 0.05 \right)$.
For polydisperse systems, we implement a swap Monte Carlo methodology
originally developed by Misaki Ozawa for \cite{berthierZerotemperatureGlassTransition2019}.
With probability $0.2$, a trial move corresponds to a trial swap instead of a trial displacement.
If a particle with size $\sigma_i$ is selected for a swap move, a second particle with size $\sigma_j$
satisfying $\left| \sigma_j - \sigma_i \right| < 0.2$ is randomly selected as its swap partner.
We start our Monte Carlo simulations from a configuration with random initial positions,
except when initializing a crystal in the monodisperse system, in which case we start from a perfect FCC crystal
in order to accelerate the equilibration process.

When preparing polydisperse systems containing an artificial target-state slab,
we use swap Monte Carlo to equilibrate the system to the target temperature $T_\mathrm{eq}$.
We then designate a slab-shaped region, corresponding to points with $x$-coordinate satisfying $\left| x \right| > a$ for some $a > 0$,
to be outside the target-state slab.
We choose a new, off-target temperature $T_0$ for this region, and find the number density for a homogeneous system
at equilibrium at temperature $T_0$ and with virial pressure matching that of the target equilibrium system.
We stretch or compress the off-slab region along the $x$-axis such that it matches this number density,
thus reducing the pressure gradient that would otherwise exist between the slab and the non-slab region.
MD simulations are then run at the temperature and pressure corresponding to the target equilibrium system.

When preparing monodisperse systems containing an artificial target-state slab,
we choose crystal and liquid temperatures that are similar in magnitude,
such that the pressure gradient is less pronounced.
It is not possible to stretch or compress the off-slab region along the $x$-axis to reduce the pressure gradient,
as we do in the polydisperse case, as this would prevent us from guaranteeing system dimensions compatible
with a perfect FCC crystal.
We first homogeneously equilibrate the system to the target crystal state,
then conduct Monte Carlo simulations at high temperature, sampling only from the liquid region.
Once the liquid region has melted, we conduct Monte Carlo simulations at the target, lower temperature for the liquid region,
sampling only from this region.
MD simulations in this case are run at constant temperature and
a constant pressure chosen such that the virial pressure at equilibrium
matches the mean virial pressure in the initial configuration.

\subsection{Molecular dynamics \label{app:MD}}

We conduct our MD simulations in LAMMPS~\cite{plimptonFastParallelAlgorithms1995} (23 Jun 2022 - Update 2)
at constant temperature and pressure implemented via LAMMPS' Nos\'e-Hoover thermostat (\texttt{fix nvt}) with damping parameter \texttt{Tdamp = 1}
and Berendsen barostat (\texttt{fix press/berendsen}) with damping parameter \texttt{Pdamp = 1}.
If a system has been initialized to temperature $T_0$ using Monte Carlo,
we find the virial contribution to the pressure at this temperature, and choose a pressure such that this contribution is
maintained once the temperature changed to the target temperature $T_\mathrm{eq}$.
We do this so as to minimize changes in the system size upon changing the temperature.

\subsection{Initial and steady-state distributions \label{app:t0_tinfty}}

Because the start-up of the thermostat and barostat perturbs the system at early times,
we must wait until a time $t_0 > 0$ before we can measure initial distributions
for the homogeneous simulations.
We use $t_0=55$ for simulations of the monodisperse system
and $t_0=100$ for the polydisperse system.
For homogeneous melting simulations in the monodisperse system,
equilibrium distributions are obtained from $t=600$ data,
restricting to realizations with bulk average $\overline\Theta \left( t=600 \right) > 0.091$.
For the homogeneous freezing simulations,
``equilibrium'' distributions are obtained from $t=10000$ data.
This does not correspond to equilibrium at the target temperature $T_\mathrm{low}$,
but rather a collection of metastable states that each realization is trapped in at that time
(see \S\ref{sec:discussion}).
For the polydisperse system, equilibrium distributions at temperature $T_\mathrm{eq} = T_\mathrm{high}$ 
are obtained from homogeneous heating simulation data for $t \geq 80000$,
while equilibrium distributions at temperature $T_\mathrm{eq}=T_\mathrm{low}$
are obtained from simulations at that temperature in the micro-canonical ensemble.

\begin{figure}
\includegraphics[width=\columnwidth]{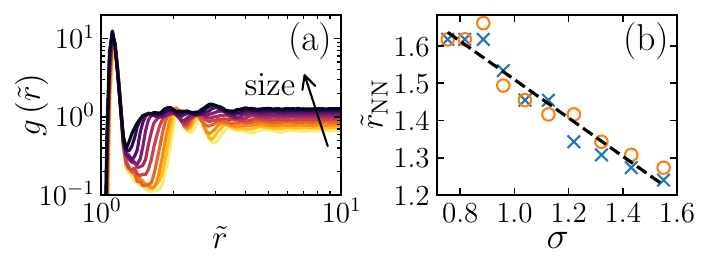}
\caption{Subplot (a): pair distribution $g \left( \tilde r \right)$
for the polydisperse system at temperature $T=0.14$,
calculated separately for particles at the origin of size $\sigma$
within each of ten logarithmically-spaced bins, plotted as separate curves (larger $\sigma$, darker curve).
This pair distribution is calculated using the same normalization as the usual pair distribution $g \left( r \right)$,
but replacing the raw interparticle separation $r$ with $\tilde r$,
hence why $g \left( \tilde r \right) \nrightarrow 1$ as $\tilde r \rightarrow \infty$.
Subplot (b): $\tilde r$-position $\tilde r_\mathrm{NN}$ of the trough
following the first nearest neighbor peak of $g \left( \tilde r \right)$
as a function of the diameter $\sigma$ of the particle at the origin for temperatures
$T=0.06$ (blue crosses) and $T=0.14$ (orange circles).
The dashed line is the linear fit yielding Eq.~\ref{eq:rNN}.
\label{fig:rNN}}
\end{figure}

\subsection{$\Theta$ \label{app:Theta}}

The main structural indicator used in this work is Tong and Tanaka's $\Theta$ order parameter~\cite{tongRevealingHiddenStructural2018},
which measures how inefficiently packed the structure local to each particle is.
For each particle $i$ in the system, we iterate over pairs of nearest neighbors $j$ and $k$
such that $j$ and $k$ are also nearest neighbors of each other.
We consider them perfectly efficiently packed if the pair potentials $V_{ij}$, $V_{jk}$, and $V_{ik}$
are all equal.
If these particles have corresponding sizes $\sigma_i$, $\sigma_j$ and $\sigma_k$,
this means their centers form a triangle with sides of length $l_{ij} = \frac{1}{2} \left( \sigma_i + \sigma_j \right)$,
$l_{ik} = \frac{1}{2} \left( \sigma_i + \sigma_k \right)$ and $l_{jk} = \frac{1}{2} \left( \sigma_j + \sigma_k \right)$.
At $i$, this triangle makes an angle $\theta_{ijk}^\ast$ such that
\begin{equation}
    \cos \theta_{ijk}^\ast = \frac{l_{ij}^2 + l_{ik}^2 - l_{jk}^2}{2 l_{ij} l_{ik}}.
\end{equation}
Letting $\theta_{ijk}$ be defined such that
\begin{equation}
    \cos \theta_{ijk} = \frac{r_{ij}^2 + r_{ik}^2 - r_{jk}^2}{2 r_{ij} r_{ik}},
    \label{eq:theta}
\end{equation}
where $r_{ij}$ is the distance separating particles $i$ and $j$, with $r_{jk}$ and $r_{ik}$ defined similarly,
we define $\Theta$ at $i$ to be
\begin{equation}
    \Theta_i = \left< \left| \theta_{ijk} - \theta_{ijk}^\ast \right| \right>_{j, k},
    \label{eq:Theta}
\end{equation}
where the average is over nearest-neighbor pairs $j$ and $k$ that are nearest neighbors of $i$ as well as each other.
The difference in angle $\left| \theta_{ijk} - \theta_{ijk}^\ast \right|$ is calculated modulo $\pi$.

For the purposes of calculating $\Theta_i$ in our two-dimensional polydisperse system,
we consider $i$ and $j$ to be nearest neighbors if they have normalized separation
$\tilde r_{ij} < \tilde r_\mathrm{NN} \left( \sigma_i \right)$, where
\begin{equation}
    \tilde r_\mathrm{NN} \left( \sigma_i \right) = 2.025 - 0.515 \sigma_i.
    \label{eq:rNN}
\end{equation}
This corresponds to the first minimum after the nearest-neighbor peak of $g \left( \tilde r \right)$,
where $g \left( \tilde r \right)$ is proportional to the probability of finding a particle
at normalized separation $\tilde r$ given a particle at the origin with diameter $\sigma$, as shown in Fig.~\ref{fig:rNN}.
For our three-dimensional monodisperse system, the corresponding quantity is $\tilde r = 1.3$.
When a particle $i$ does not have any pair of nearest neighbors $j$ and $k$ that are nearest neighbors of each other,
we leave its $\Theta_i$ undefined and exclude it when calculating averaged quantities.

\subsection{Mermin-Wagner fluctuations \label{app:MW}}

Our two-dimensional glass is influenced by Mermin-Wagner fluctuations%
~\cite{flennerFundamentalDifferencesGlassy2015, vivekLongwavelengthFluctuationsGlass2017, illingMerminWagnerFluctuations2017}.
We account for this by subtracting from any displacement vector $\Delta p_i$ of a particle $i$
the mean displacement vector $\left< \Delta p_j \right>_j$ averaged over neighboring particles $j \neq i$
within a distance $\xi$ of particle $i$.
We choose the value of $\xi$ to be that appropriate to equilibrium at temperature $T_\mathrm{eq}$,
determined as in \cite{chackoElastoplasticityMediatesDynamical2021}.

\subsection{FCC Ground State \label{app:FCC_ground}}

\begin{figure}
\includegraphics[width=.5\columnwidth]{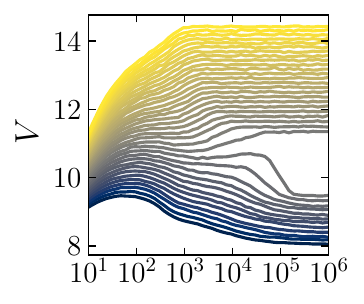}
\caption{Evolution in the canonical ensemble of the mean potential energy of an initially perfect HCP crystal
with $22$ particles along each basis vector.
Curves correspond to temperature $T=1.5$, $1.6$, $\dots$, $5.0$ (lighter color, higher temperature).
Time is in units of number of trial Monte Carlo steps per particle.
\label{fig:FCC_ground_state}}
\end{figure}

To determine that the ground state of the monodisperse system is an FCC crystal,
we reason that the ground state of an isotropic repulsive pair interaction decreasing monotonically with distance
must correspond to the distance-maximizing FCC or HCP (hexagonal close packing) structures.
As shown in Fig.~\ref{fig:FCC_ground_state}, HCP packings at temperature $T < 2.7$
(thermostat implemented using Monte Carlo trial steps at constant volume)
transiently increase in potential energy before relaxing into what we find to be an FCC crystal,
implying that the FCC state is more stable.
(For $T \geq 2.7$, the packings melt into a liquid instead.)

\subsection{Interpolating distributions \label{app:distbn_interp}}

\begin{figure}
\includegraphics[width=\columnwidth]{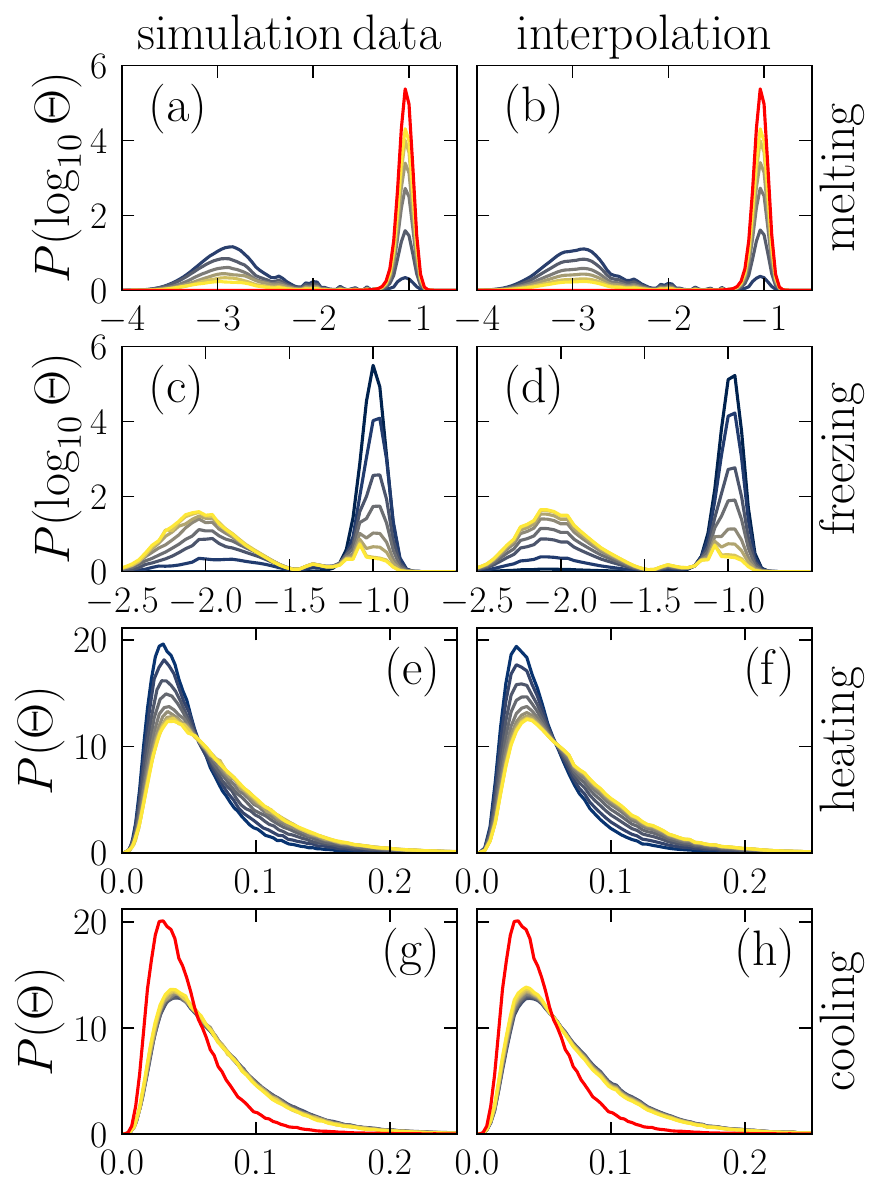}
\caption{Validation of Eq.~\ref{eq:avrami}.
Each row corresponds to data from the main text:
(a, b): Fig.~\ref{fig:homog_crystal}(b),
(c, d): Fig.~\ref{fig:homog_crystal}(f),
(e, f): Fig.~\ref{fig:homog_poly}(b),
(g, h): Fig.~\ref{fig:homog_poly}(f).
Left hand side: same data as in the corresponding figure from the main text.
Right hand side: output of Eq.~\ref{eq:avrami}
(where $\phi = \log_{10} \Theta$ (a--d) or $\Theta$ (e--h))
with $f$ calculated from Eq.~\ref{eq:mean} applied to the quantity $\Theta$.
\label{fig:distbn_interp}}
\end{figure}

In the main text, we calculate the transformed fraction $f$ from the evolving distribution $P \left( \phi, t \right)$
for different quantities $\phi$ characterizing the local structure or mobility of the system,
plotting them as dashed curves in Figs.~\ref{fig:homog_crystal}(d), \ref{fig:homog_crystal}(h),
\ref{fig:homog_poly}(d), and \ref{fig:homog_poly}(h).
We achieve this using one of two approaches, depending on the behavior of the growing $P_\mathrm{eq} \left( \phi \right)$ mode.
When fluctuations in the distribution near the $P_\mathrm{eq} \left( \phi \right)$ mode are not too severe,
we use
\begin{equation}
    f (t) = \frac{P \left( \phi_\ast, t \right) - P_0 \left( \phi_\ast \right)}{P_\mathrm{eq} \left( \phi_\ast \right) - P_0 \left( \phi_\ast \right)},
    \label{eq:interp_Theta}
\end{equation}
where $\phi_\ast = \argmax\limits_{\phi} P_\mathrm{eq} \left( \phi \right)$.
This is the case for $1-f$ calculated from $P \left( \log_{10} \Theta \right)$ and
$P \left( \log_{10} \mu \right)$ in Fig.~\ref{fig:homog_crystal}(d)
(dashed red curve and dashed green curve, respectively),
and for $1-f$ calculated from $P \left( \Theta \right)$ in
Figs.~\ref{fig:homog_poly}(d) and \ref{fig:homog_poly}(h) (dashed red curves).
However, where the growing $P_\mathrm{eq} \left( \phi \right)$ mode is sensitive to noise,
or where it shifts slightly as it grows (Fig.~\ref{fig:homog_poly}(c)),
we use
\begin{equation}
    f (t) = \frac{\max\limits_{\phi \in I} P \left( \phi, t \right) - \max\limits_{\phi \in I} P_0 \left( \phi \right)}{\max\limits_\phi P_\mathrm{eq} \left( \phi_\ast \right) - \max\limits_{\phi \in I} P_0 \left( \phi \right)},
    \label{eq:interp_using_interval}
\end{equation}
which is insensitive to these fluctuations.
Here, $I$ is an appropriately-chosen interval.
This approach is used to obtain the red and green dashed curves in Fig.~\ref{fig:homog_crystal}(d)
($I = \left( - \infty, -1.75 \right)$ (red) and $\left( - \infty, -2 \right)$ (green)),
as well as the green dashed curve in Fig.~\ref{fig:homog_poly}(d)
($I = \left( -0.5, \infty \right)$).

We also calculate $f$ from Eq.~\ref{eq:mean},
using the averages $\overline\phi$ of various quantities $\phi$ and plotting them as solid or dotted curves
in Figs.~\ref{fig:homog_crystal}(d), \ref{fig:homog_crystal}(h), \ref{fig:homog_poly}(d), and \ref{fig:homog_poly}(h).
To demonstrate that our different methods of calculating $f$, which collapse onto one another,
successfully yield the interpolation parameter $f$ in Eq.~\ref{eq:avrami},
we compare in Fig.~\ref{fig:distbn_interp} the evolutions $P \left( \phi, t \right)$ shown in
Figs.~\ref{fig:homog_crystal}(d), \ref{fig:homog_crystal}(h), \ref{fig:homog_poly}(d), and \ref{fig:homog_poly}(h)
to the evolution obtained from Eq.~\ref{eq:avrami} where $P_0 \left( \phi \right)$ and $P_\mathrm{eq} \left( \phi \right)$
are taken as input from simulation data and $f$ is obtained from Eq.~\ref{eq:mean}
using $\phi = \log_{10} \Theta$ (for the crystal) or $\Theta$ (for the glass).

\section{Supporting Details}

\subsection{Perturbed Crystal \label{app:perturbed}}

\begin{figure}
\includegraphics[width=\columnwidth]{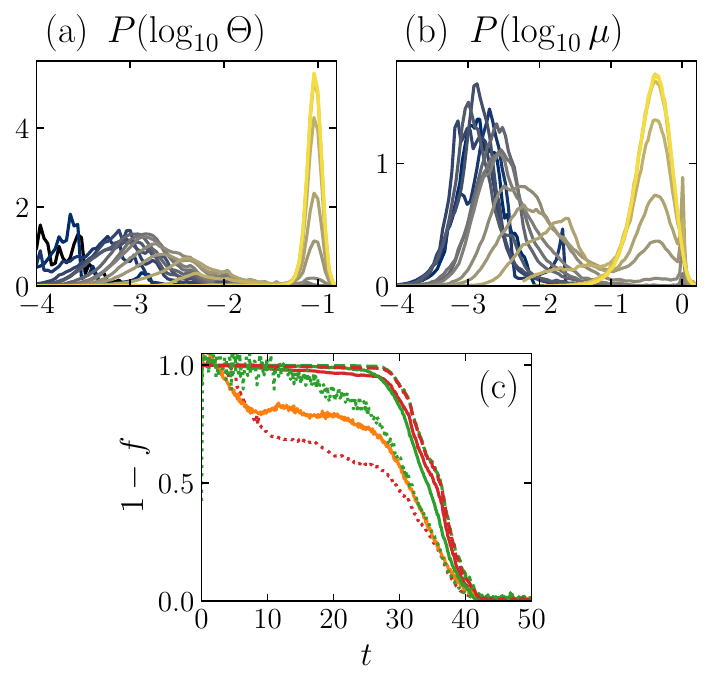}
\caption{Melting transition in a monodisperse system heated from $T_0=1.90$
         to $T_\mathrm{eq}=2.40$
         (\emph{c.f.} $T_\mathrm{eq}=2.35$ in Figs.~\ref{fig:homog_crystal}(a)--\ref{fig:homog_crystal}(d)).
         Distributions (a) $P \left( \log_{10} \Theta \right)$
         and (b) $P \left( \log_{10} \mu \right)$
         at times $t=0.0$, $3.2$, $\dots$, $48.0$
         (later time, lighter curve).
         (c): untransformed fraction $1-f$ for $f$ calculated from
         (dashed red) $P \left( \log_{10} \Theta \right)$
         and (dashed green) $P \left( \log_{10} \mu \right)$,
         plotted against the mean
         (solid orange) potential energy $V$,
         (solid red) $\Theta$, (dotted red) $\log_{10} \Theta$,
         (solid green) $\mu$, and (dotted green) $\log_{10} \mu$,
         affinely transformed according to Eq.~\ref{eq:mean}
         with $t_0 = 2.0$.
         On these short time scales, start-up of the barostat leads to
         erratic behavior of the inherent state energy, which is therefore not shown.
\label{fig:homog_crystal_high_T}}
\end{figure}

In \S\ref{sec:crystal} and \S\ref{sec:glasses}, we considered systems well-described by
the simple interpolation between initial and target states expressed in Eq.~\ref{eq:avrami}.
We stressed that this implies a clean two-state equilibration scenario,
as the systems evolve directly from the initial state into the final state.
A good test of this idea is provided by systems in which the initial state is perturbed
during the heating process, violating Eq.~\ref{eq:avrami}.
In Fig.~\ref{fig:homog_poly}, for instance, there is a visible rightwards shift of
the $P_0 \left( \log_{10} \mu \right)$ peak in Fig.~\ref{fig:homog_poly}(c) as
target-state domains grow and a corresponding deviation of the decay of the mean of $\log_{10} \mu$
from that of the fraction $f$ of un-transformed structure in Fig.~\ref{fig:homog_poly}(d).

Here, we show in Fig.~\ref{fig:homog_crystal_high_T} that slightly increasing the temperature
at which we heat our crystal ($T_\mathrm{high}=1.40$,
as compared to $T_\mathrm{high}=1.35$ for melting in Fig.~\ref{fig:homog_crystal})
leads to large rightwards shifts of the $P_0 \left( \log_{10} \Theta \right)$
and $P_0 \left( \log_{10} \mu \right)$ peaks as the melting proceeds
(Figs.~\ref{fig:homog_crystal_high_T}(a) and \ref{fig:homog_crystal_high_T}(b)),
with corresponding deviations of $1-f$ calculated from Eq.~\ref{eq:mean}
using the mean values of $\log_{10} \Theta$ and $\log_{10} \mu$
from the correct value of $1-f$ obtained from the heights of the 
$P_\mathrm{eq} \left( \log_{10} \Theta \right)$ and $P_\mathrm{eq} \left( \log_{10} \mu \right)$ peaks
(Fig.~\ref{fig:homog_crystal_high_T}(c)).
Inspection of individual trajectories (not shown) suggests that this deviation is due to a large defect density,
with pairs of defects imposing strain on crystalline domains trapped between them
(note the visible interface of length $\approx 4$ particle diameters at the edge of the growing domain
in Fig.~\ref{fig:homog_crystal}(a), top row).

By contrast, we find that increasing $T_\mathrm{high}$ to $0.17$ in the homogeneous heating of the polydisperse glass,
far above $T_\mathrm{onset}$,
does not lead to a perturbation of the un-transformed structure
and resultant violation of the two-state equilibration scenario
beyond that seen in Fig.~\ref{fig:homog_poly} for $\log_{10} \mu$.
This is likely the result of the highly localized nature of facilitation at low temperatures,
as seen in Fig.~3(d) of \cite{chackoElastoplasticityMediatesDynamical2021},
and the fact that domains of $T_\mathrm{high}$ and $T_\mathrm{low}$ do not noticeably statically interact,
as we establish in the main text.

\subsection{Two-state scenario test: three-state case \label{app:twoStateScenario}}

Our two-state scenario test allows us to distinguish multi-step relaxation processes
from the direct transformation of material from initial-state structure
into target-state structure.
To illustrate this, we consider a system that fully transforms from the initial state
into an intermediate state before transforming into the target equilibrium state.
In this case, we have
\begin{equation}
    \overline\phi = \left( 1 - f_0 \right) \overline\phi_0 + \left( f_0 - f_1 \right) \overline\phi_1
    + f_1 \overline\phi_\mathrm{eq},
    \label{eq:mean_twostep}
\end{equation}
where $f_0$ is the fraction of the system that has transformed from the initial state into the intermediate state,
$f_1$ (which vanishes unless $f_0 = 1$)
is the fraction of the system that has transformed from the intermediate state into the target state,
and $\overline\phi_1$ is the mean of $\phi$ in the intermediate state.
It is clear from Eq.~\ref{eq:mean_twostep} that in this case,
$1-f$ calculated according using Eq.~\ref{eq:mean} will be different for different quantities $\phi$.

\begin{figure}
\includegraphics[width=\columnwidth]{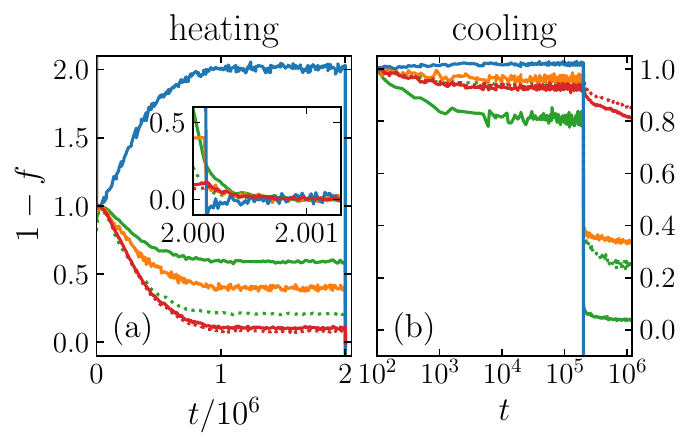}
\caption{
(a) Heating from temperature $T_0 = 0.06$ to $T_1 = 0.12$ until time $t = 2 \times 10^6$,
followed by heating at $T_\mathrm{eq} = 0.14$.
(b) Cooling from temperature $T_0 = 0.14$ to $T_1 = 0.12$ until time $t = 2 \times 10^5$,
followed by cooling at $T_\mathrm{eq} = 0.06$.
Plots show $1-f$ calculated from
potential energy $\overline V$ (solid orange),
inherent state potential energy $\overline\VIS$ (solid blue),
$\overline\Theta$ (solid red),
$\overline{\log_{10} \Theta}$ (dotted red),
$\overline{\mu}$ (solid green),
and $\overline{\log_{10} \mu}$ (dotted green).
Inset to (a): zoom into times during which $1-f$ evolves rapidly.
\label{fig:twostate_twostep}}
\end{figure}

To confirm this, we adapt the homogeneous simulations of \S\ref{sec:glasses},
introducing an intermediate temperature $T_1 = 0.12$.
We evolve the glass from its initial $T_0$ equilibrium state until steady state
with a thermostat at temperature $T_1 = 0.12$,
then change the thermostat temperature to $T_\mathrm{eq}$ and evolve the system further.
We fix the pressure based on $T_\mathrm{eq}$,
matching the pressures used in the homogeneous simulations of \S\ref{sec:glasses}.

As shown in Fig.~\ref{fig:twostate_twostep}, the curve collapses seen in
Figs.~\ref{fig:homog_poly}(d) and \ref{fig:homog_poly}(d)(h)
are not recovered in this multi-stage relaxation scenario.
For both heating and cooling, the structural quantities
$V$, $\VIS$, $\Theta$, and (to a lesser extent) $\log_{10} \Theta$
evolve differently to one another,
both during the initial evolution at the intermediate temperature $T_1 = 0.12$,
and after the thermostat is changed to the target temperature $T_\mathrm{eq}$.
This validates calculating $1-f$ from Eq.~\ref{eq:mean} for different structural quantities $\phi$
as a test of the two-state scenario.

\subsection{Structural perturbation due to rearrangements \label{app:struct_perturb}}

\begin{figure}
\includegraphics[width=\columnwidth]{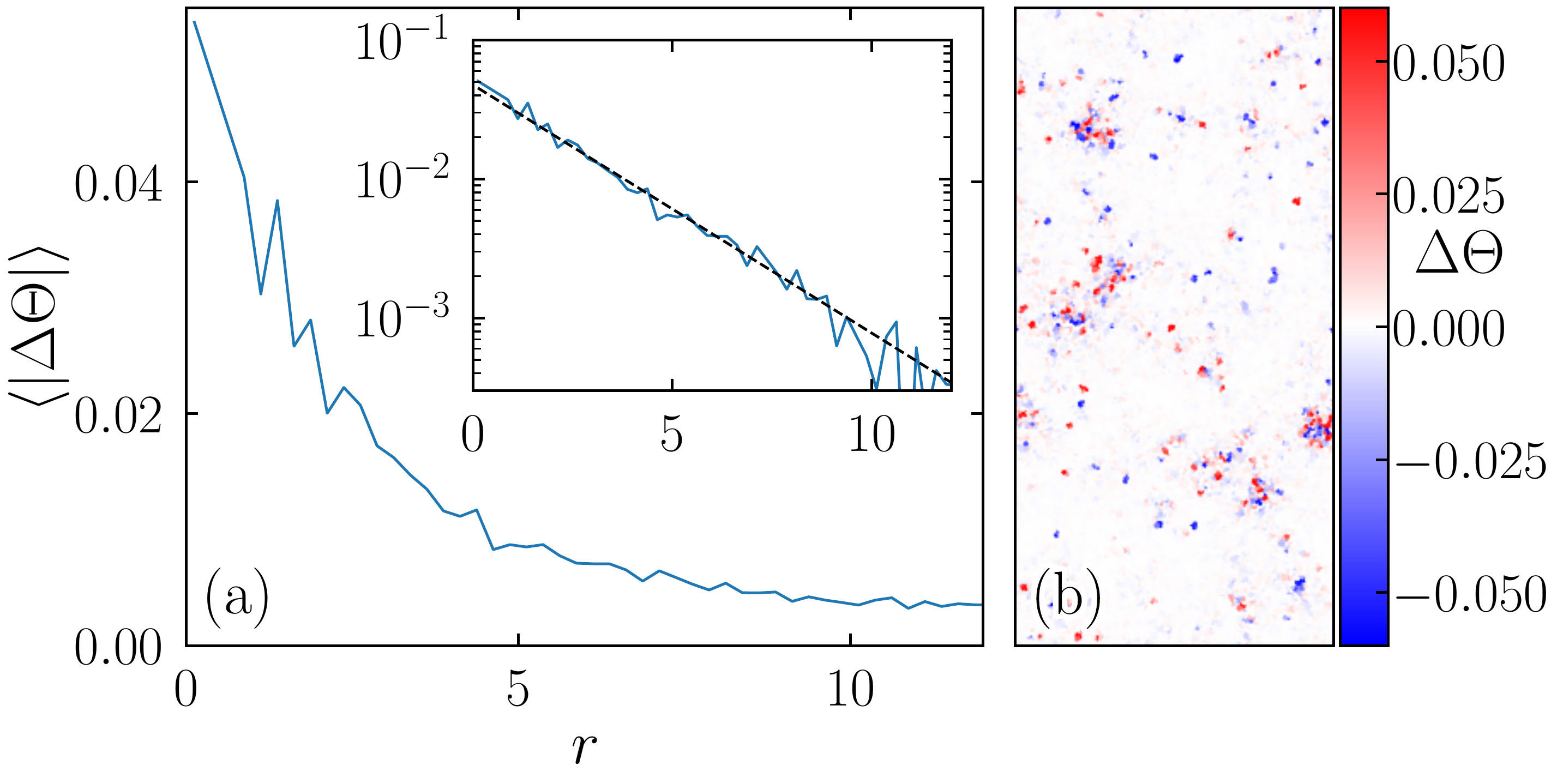}
\caption{Perturbation to the local $\Theta$ field due to rearrangements
in the homogeneously cooled glass studied in Figs.~\ref{fig:homog_poly}(e)--\ref{fig:homog_poly}(h)
at time $t=10^6$.
(a): Mean absolute change in $\Theta$ between times $10^6$ and $10^6 + 10^2$ for particles a distance $r$ from a rearranging particle,
averaged over ten independent realizations.
The inset shows this data on a log-scale after subtracting the global mean absolute change in $\Theta$.
The dashed line corresponds to a fit to an exponential decay with decay length $\xi_\mathrm{perturb} = 2.43$.
(b): Change in $\Theta$ across the same time interval as (a) for the realization shown in Fig.~\ref{fig:homog_poly}(e).
\label{fig:abs_dTheta}}
\end{figure}

As the glass evolves during cooling in Fig.~\ref{fig:homog_poly}(e),
it becomes possible to distinguish individual rearrangement events from the characteristic quadrupolar displacement fields they induce,
as seen in the snapshots at times $t \geq 10^3$.
In Fig.~\ref{fig:abs_dTheta}, we leverage this to identify the effect of such individual rearrangements on the local structure,
identifying rearrangements with particles that have mobility $\mu > 0.1$
and have a larger mobility than any other particle within $8$ mean particle diameters of distance.
We see that changes in $\Theta$ near the rearrangement are of the same magnitude $\sim 10^{-1}$ as $\Theta$ itself
(Figs.~\ref{fig:homog_poly}(b) and \ref{fig:homog_poly}(b))
up to a distance $r \approx 5$ from the rearrangement,
decaying exponentially with decay length $\xi_\mathrm{perturb} = 2.34$.

\subsection{Local relaxation in the TPM \label{app:TPM_fit}}

\begin{figure}
\includegraphics[width=\columnwidth]{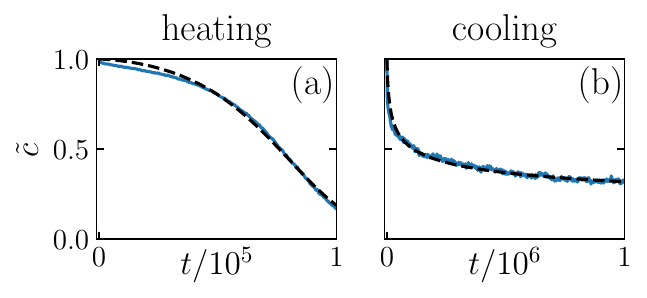}
\caption{Fits (dashed lines) to the decay of $\tilde c \left( x = \pm 64 \right)$
         towards equilibrium in the TPM during (a) heating (\ref{fig:inhom_plaquette}(e))
         and (b) cooling (\ref{fig:inhom_plaquette}(g)).
         In these plots, $\tilde c$ is transformed according to Eq.~\ref{eq:mean}
         to make the decay clear.
\label{fig:PDE_fit}}
\end{figure}

Appropriate values for parameters $\tau_0$ and $a$ in Eq.~\ref{eq:tau} are obtained by fitting
Eq.~\ref{eq:PDE} for a homogeneous system to the decay of $\tilde c \left( x = \pm 64 \right)$
towards equilibrium in Figs.~\ref{fig:inhom_plaquette}(e) and \ref{fig:inhom_plaquette}(g).
We show these fits in Fig.~\ref{fig:PDE_fit}.

\subsection{Supplemental Movies \label{app:supp_movies}}

In \footnote{See Supplemental Material for six supplemental movies.},
we include supplemental movies for the homogeneous simulations found in the main text,
including many more snapshots than could be put in the main text.
We show multiple realizations of each simulation,
with the realization in the main text included as the left-most column.
We pair the snapshots with the corresponding evolution of $1-f$ for the given realization,
calculated as in Fig.~\ref{fig:decays}.
Supplemental Movie 1 corresponds to Figs.~\ref{fig:homog_crystal}(a) and \ref{fig:decays}(a),
Supplemental Movie 2 corresponds to Figs.~\ref{fig:homog_crystal}(e) and \ref{fig:decays}(b),
Supplemental Movie 3 corresponds to Figs.~\ref{fig:homog_poly}(a) and \ref{fig:decays}(c),
and Supplemental Movie 4 corresponds to Figs.~\ref{fig:homog_poly}(e) and \ref{fig:decays}(d).
Supplemental Movies 5 and 6 are copies of Supplemental Movies 3 and 4, respectively,
but with the $\Theta$ plot replaced with the change $\Delta \Theta$ in $\Theta$
between times $t$ and $t+100$.

\newpage

\bibliography{library}

\end{document}